\documentclass[universe,article,accept,pdftex,oneauthor]{Definitions/mdpi}  

\firstpage{1} 
\pubvolume{0}
\issuenum{0}
\articlenumber{0}
\pubyear{2026}
\copyrightyear{2026}
\externaleditor{Academic Editor: ...}
\datereceived{...} 
\dateaccepted{...} 
\datepublished{...}

\hreflink{https://\linebreak 
	doi.org/...} 
\doinum{...}

\usepackage{bm}
\usepackage{amsmath,amssymb} 
\usepackage{wasysym}
\usepackage{pifont,bbm}
\usepackage{stmaryrd}
\usepackage{graphbox,graphicx}
\usepackage{arydshln}
\usepackage{multirow}
\newcommand{\vvarpi}{\varpi}
\newcommand{\xxx}{y_1}
\newcommand{\yyy}{y_2} 
\newcommand{\ktilde}{\tilde\kappa}
\newcommand{\rin}{\vvarpi_{\mbox{\tiny CD}}}

\newcommand{\rhoin}{\rho_{\mbox{\tiny CD}}}
\newcommand{\zin}{{\rm z}_{\mbox{\tiny CD}}}
\newcommand{\F}{F}
\newcommand{\W}{W}

\newcommand{\apj}{Astroph. J.}
\newcommand{\apjs}{Astroph. Space Sci.}
\newcommand{\apjl}{Astroph. J.}
\newcommand{\mnras}{Mon. Not. R. Astron. Soc.}
\newcommand{\aap}{Astron. Astrophys.}
\newcommand{\bmath}{\bm}
\newcommand{\be}{\begin{equation}}
\newcommand{\ee}{\end{equation}}

\makeatletter
\renewcommand{\maketag@@@}[1]{\hbox{\m@th\normalsize\normalfont#1}}
\makeatother

\Title{Instabilities in Cylindrical Geometry Using the Minimalist Approach: Formalism and Rotational Instabilities}

\TitleCitation{Instabilities in Cylindrical Geometry Using the Minimalist Approach: Formalism and Rotational Instabilities}

\Author{Nektarios Vlahakis \orcidA{}}

\AuthorNames{Nektarios Vlahakis}

\AuthorCitation{Vlahakis, N.}

\address[1]{%
Department of Physics, National and Kapodistrian University of Athens, University Campus,\linebreak Zografos GR-157 84 Athens, Greece; vlahakis@phys.uoa.gr
} 

\abstract{The minimalist approach for linear stability analysis is applied to fluids and magnetized ideal plasmas in cylindrical geometry. In this approach, the dispersion relation is obtained by integrating a single first-order differential equation -- referred to as the principal equation -- subject to appropriate boundary conditions.
We first derive the principal equation for a general unperturbed state with radially varying density and pressure, axial and azimuthal components of both the velocity and magnetic field, and a radially directed gravitational field.
We then use this formulation to analyze rotating flows with axial magnetic fields, addressing both wall-bounded and interface-driven axisymmetric instabilities. In addition to exact results for selected unperturbed states, we obtain approximate dispersion relations using the WKBJ method in the incompressible and compressible limits.
The analysis encompasses centrifugal, magnetorotational, and buoyancy-driven instabilities as special cases, and it clarifies how compressibility modifies their stability properties.}

\keyword{instabilities; fluid dynamics; hydrodynamics; magnetohydrodynamics; analytical methods} 

\begin{document}

\section{Introduction}\label{introduction}

Cylindrical flows play a foundational role in astrophysics. Their geometry arises naturally from rotation and provides an effective framework for describing matter and magnetic field behavior in extreme environments.
Astrophysical jets offer a clear example. Jets from active galactic nuclei, microquasars, gamma-ray bursts, and protostars maintain narrow, elongated shapes over vast distances. Modeling them as cylindrical plasma beams captures their essential dynamics: rotation winds magnetic fields into helical structures that confine and guide the jet.
Accretion disks form a second major domain where cylindrical geometry is indispensable. Around compact objects and young stars, matter spirals inward under gravity while maintaining nearly Keplerian rotation. Cylindrical coordinates highlight the dominant azimuthal motion, strong shear between neighboring fluid layers, which drives turbulence and magnetic field amplification.  

Although jets and disks operate on much larger scales, their cylindrical geometry is mirrored in laboratory plasma devices based on pinch-type configurations, rotating plasma columns, liquid-metal experiments, and laser-driven high-energy-density plasmas. These systems provide controlled analogues that help test and validate theoretical models used in astrophysics. 
Additional insight comes from other areas of fluid dynamics where rotating and stratified flows have been studied extensively, as in geophysical fluid dynamics. Despite operating in very different parameter regimes, the mathematical structure of the governing equations is closely related, and techniques developed in those contexts remain directly applicable to cylindrical astrophysical flows.

Stability is central to fluid dynamics because it determines whether a flow configuration can persist and how it evolves. Jets and disks exist in environments dominated by rotation, shear, magnetic fields, and stratification, making them sensitive to small disturbances that can bend, kink, fragment, or drive the flow into turbulence. Stability analysis predicts which outcomes are possible and under what conditions.
Stability also governs efficiency and lifetime. For example a stable jet remains collimated and transports energy over large distances, whereas an unstable jet disrupts and deposits energy locally. In this sense, stability analysis is not merely a mathematical exercise but a tool for interpreting the structure, energetics, and evolution of astrophysical flows.

Linear stability analysis is important because every instability begins as an infinitesimal perturbation. By examining the linearized equations of motion, one can determine whether such perturbations grow or decay, identify their characteristic spatial structure, and compute their growth rates. This provides a clear and predictive framework for understanding the onset of instability before nonlinear effects become dominant. Linear theory also reveals which mechanisms drive instability and it often yields results that are universal across a wide range of physical conditions. Moreover, the early stages of instability frequently leave observable signatures, such as oscillations or coherent distortions, that correspond directly to linear modes. Numerical simulations rely on linear theory to validate their behavior and to interpret the transition from linear growth to nonlinear saturation.

Rotational instabilities arise because rotation stores free energy that can be released when perturbations interact with shear, magnetic fields, or stratification. They fall into three main categories: centrifugal instabilities (CFI), magnetorotational instabilities (MRI), and buoyancy-driven instabilities.

CFI is purely hydrodynamic and occurs when the specific angular momentum decreases outward, violating the Rayleigh criterion \cite{1917RSPSA..93..148R}. Fluid elements displaced outward then lack sufficient angular momentum to remain in orbit. Rayleigh’s early analysis remains the cornerstone of rotational stability theory, and CFI can appear in strongly sheared or transitional regions of accretion disks.

MRI is the magnetohydrodynamic counterpart. A weak magnetic field couples fluid elements at different radii, allowing angular-momentum exchange and destabilizing flows that are hydrodynamically stable. The seminal work of Balbus and Hawley~\cite{1991ApJ...376..214B}, following earlier studies~\cite{1959JETP....9..995V,1960PNAS...46..253C}, established MRI as a dominant mechanism for angular-momentum transport in accretion disks around compact objects and young stars.

A distinct line of recent work by Gourgouliatos and Komissarov~\cite{2018MNRAS.475L.125G} has expanded the concept of rotational instabilities by exploring how they operate in recollimation shocks of astrophysical jets. Their studies, which have been generalized including magnetic fields, show that when a jet undergoes recollimation -- typically due to pressure mismatches with the surrounding medium -- the flow can develop regions locally susceptible to CFI/MRI. Thus rotational instabilities are not limited to pure cylindrical flows but may be triggered in dynamically evolving jet structures, providing a powerful mechanism for energy dissipation and magnetic field evolution.

Buoyancy-driven rotational instabilities arise when a stratified cylindrical flow is subject to an effective gravity composed of real gravitational acceleration plus the outward centrifugal acceleration generated by rotation. Unstable stratification leads to convection, where buoyant fluid rises and denser fluid sinks. In compressible media, stability depends on how density and pressure vary together, quantified by the buoyancy (or Brunt-V\"ais\"al\"a) frequency. In magnetized flows, buoyancy interacts with the magnetic field, producing magnetic-buoyancy (Parker) instability~\cite{1966ApJ...145..811P}, in which magnetic field lines rise under effective gravity while plasma slides downward along them.

Together, centrifugal, magnetorotational, and buoyancy-driven instabilities form the core framework for rotational stability in cylindrical astrophysical flows, revealing how rotation, magnetic fields, and stratification combine to shape the evolution of jets, disks, and plasma columns.

A standard way to obtain the dispersion relation is to start from the governing ideal MHD equations -- momentum, induction, continuity, and energy -- and linearize them around a chosen one-dimensional equilibrium. Introducing small perturbations and adopting a normal-mode form converts the linearized PDEs into a set of ODEs in the cylindrical radius for the perturbation amplitudes. From this point, several established solution strategies exist, including spectral methods as in the code Legolas~\cite{2020ApJS..251...25C}, the Spectral Web~\cite{2018PhPl...25c2109G}, and shooting techniques. In addition, closed-form dispersion relations -- rarely obtainable except under strong simplifying assumptions -- and local analytic approximations such as WKBJ analysis offer physical insight into the underlying instability mechanisms. 

In this paper we follow the minimalist approach~\cite{2024Univ...10..183V}, in which one eliminates variables systematically until the system reduces to a single first-order ODE, the principal equation. The dispersion relation then follows from the requirement that this equation admit non-trivial solutions satisfying the appropriate boundary conditions. 

A particularly effective way to integrate the principal equation, especially when the eigenfunctions are real, is the Schwarzian approach~\cite{2024Symm...16..648V}, which naturally numbers the eigenmodes providing a clean organizational structure for the spectrum.    

The methodology presented here provides a compact and unified framework for global stability analysis in cylindrical geometry. It shows how centrifugal, magnetorotational, and buoyancy-related modes -- incompressible and compressible -- can be treated within a single coherent formalism, with the principal equation, boundary condition treatment, and Schwarzian approach offering a streamlined alternative to existing analyses.

The plan of the paper is as follows. In Section~\ref{seclinearanalysis} we derive the principal equation in cylindrical geometry, outline the relevant boundary conditions, and describe how the Schwarzian approach can be used to find the eigenmodes. We then apply the method to specific examples, obtaining the exact dispersion relation either through numerical integration or through closed-form expressions, and we also derive an approximate dispersion relation using the WKBJ method. Section~\ref{secwall-b} treats a wall-bounded incompressible rotating flow, while Section~\ref{secinterface} examines an incompressible rotating flow with a tangential discontinuity as a representative interface-driven instability. In Section~\ref{seccompres} we extend these examples to the compressible regime to assess how compressibility modifies the instability properties. Section~\ref{secsummary} offers a concise overview of the results.

\section{Linear Analysis, the Principal Equation, and Boundary Conditions}\label{seclinearanalysis}

The non-relativistic ideal MHD equations in Lorentz-Heaviside units are:
\begin{eqnarray}
	\left( \dfrac{\partial}{\partial t} + {\bmath{V}} \cdot
	\nabla \right)\rho=-\rho \nabla\cdot\bm V  \,, \label{mass}
	\\
	\left( \dfrac{\partial}{\partial t} + {\bmath{V}} \cdot
	\nabla \right)P=c_s^2\left( \dfrac{\partial}{\partial t} + {\bmath{V}} \cdot
	\nabla \right)\rho \,, \label{energy}
	\\
	\rho
	\left( \dfrac{\partial}{\partial t} + {\bmath{V}} \cdot
	\nabla \right)
	{\bmath{V}}=
	-\nabla\left(P+\dfrac{B^2}{2}\right)+(\bm B\cdot\nabla)\bm B+\rho\bm g  \,, \label{momentum}
	\label{mom1}
	\\
	\left( \dfrac{\partial}{\partial t} + {\bmath{V}} \cdot
	\nabla \right)\bm B=(\bm B\cdot\nabla)\bm V- \bm B(\nabla\cdot\bm V) 
	\,, \label{induction}
	\\ 
	\nabla \cdot \bmath B=0 \label{divB}  \,. 
\end{eqnarray}  
We are interested in exploring the perturbations of a steady-state that depends only on the cylindrical radius $\vvarpi$. In particular, we assume that the unperturbed state has density $\rho_0(\vvarpi)$, pressure $P(\vvarpi)$, bulk velocity $\bm V_0=V_{0z}(\vvarpi)\hat z + V_{0\phi}(\vvarpi)\hat \phi$, magnetic field
$\bm B_0=B_{0z}(\vvarpi)\hat z + B_{0\phi}(\vvarpi)\hat \phi$ and
that the gravitational field is $\bm g=-g (\vvarpi)\hat\vvarpi $.
Introducing the total pressure 
\begin{eqnarray} 
	\Pi = P + \dfrac{B^2}{2} \,,
	\label{Pi1}  
\end{eqnarray} 
the zeroth-order equations are satisfied provided that  
\begin{eqnarray}
\Pi_0' =-\rho_0 g_{\rm eff}-\dfrac{B_{0\phi}^2}{\vvarpi}
\,,
\end{eqnarray} 
where the effective gravity includes the centrifugal acceleration 
\begin{eqnarray}
g_{\rm eff}=g-\dfrac{V_{0\phi}^2}{\vvarpi}
\,.
\end{eqnarray} 
(Note that $g_{\rm eff}$ may be negative when the centrifugal acceleration dominates, while the corresponding vector acceleration is $-g_{\rm eff}\hat\vvarpi$.) 

Adding perturbations of the form 
\begin{eqnarray}
	\bmath V =\bm V_0
+\left[V_{1z}(\vvarpi)\hat z+ V_{1\vvarpi}(\vvarpi)\hat \vvarpi + V_{1\phi}(\vvarpi)\hat \phi\right] 
e^{i(m\phi+kz-\omega t)} \,, 
\\
\bmath B =\bm B_0
+\left[B_{1z}(\vvarpi)\hat z+ B_{1\vvarpi}(\vvarpi)\hat \vvarpi + B_{1\phi}(\vvarpi)\hat \phi\right]e^{i(m\phi+kz-\omega t)}\,, 
\\
\rho =\rho_0(\vvarpi)
+ \rho_1(\vvarpi)e^{i(m\phi+kz-\omega t)} \,,
\\
\Pi =\Pi_0(\vvarpi)
+ \Pi_1(\vvarpi)e^{i(m\phi+kz-\omega t)}\,,
\end{eqnarray}  
with integer $m$,
defining the wavevector on the cylindrical surfaces
\begin{eqnarray}\bm k_0=k\hat z+\dfrac{m}{\vvarpi}\hat\phi \,, \end{eqnarray}
and the Doppler-shifted frequency 
\begin{eqnarray}\omega_0
=\omega-kV_{0z}-\dfrac{m}{\vvarpi}V_{0\phi}
\,, \end{eqnarray}
we linearize the equations in the Eulerian approach following a procedure similar to Ref.~\cite{1992A&A...256..354A}, shown in Appendix~\ref{appendixA}. 
The same equations can be obtained using the Frieman-Rotenberg formulation \cite{FR60} following a Lagrangian approach, as shown in Appendix~\ref{appendix_frieman}. 

The linearized system reduces to two differential equations for the perturbations of the velocity in the $\hat \vvarpi$ direction and the total pressure (there are algebraic relations for all the other perturbations, connecting them to $V_{1\vvarpi}$, $\Pi_1$, and their derivatives). 

Instead of $V_{1\vvarpi}$ it is convenient to use $\xxx=\dfrac{\vvarpi V_{1\vvarpi}}{-i\omega_0}$, a quantity connected to the Lagrangian displacement in the $\hat \vvarpi$ direction.\endnote{The relation between the Lagrangian displacement $\bm\xi$ and the velocity perturbation results
from the expression of the velocity in the perturbed location of each fluid element $\bm V+(\bm \xi\cdot\nabla)\bm V\approx \bm V_0+\delta\bm V+(\bm \xi\cdot\nabla)\bm V_0$ which equals $\bm V_0+\dfrac{d\bm\xi}{dt}\approx \bm V_0+\dfrac{\partial\bm\xi}{\partial t}+(\bm V_0\cdot\nabla)\bm \xi$, yielding $\delta\bm V=\dfrac{\partial\bm\xi}{\partial t}+(\bm V_0\cdot\nabla)\bm \xi-(\bm \xi\cdot\nabla)\bm V_0
=-i\omega_0\bm\xi-\xi_\vvarpi V_{0z}'\hat z-\xi_\vvarpi \vvarpi \left(\dfrac{V_{0\phi}}{\vvarpi}\right)'\hat \phi $.
The $\hat \vvarpi$ component gives $\xi_\vvarpi=\dfrac{\xxx(\vvarpi)}{\vvarpi} e^{i(m\phi+k z-\omega t)}$ with $V_{1\vvarpi}=-i\omega_0\dfrac{\xxx}{\vvarpi}$.}

In addition, instead of $\Pi_1$ it is convenient to use the perturbation of the total pressure evaluated at the perturbed location of each fluid element $\yyy=\Pi_1+\dfrac{\xxx}{\vvarpi}\Pi_0' $. 

The advantage of these replacements is that the new functions $\xxx$ and $\yyy$ are everywhere continuous, even at locations where the unperturbed state has tangential discontinuities.

\subsection{The System for $\xxx$, $\yyy$} 
The resulting $2\times 2$ system is
\begin{eqnarray}
	\frac{d}{d\vvarpi} \left( \begin{array}{c}
	y_1 \\ y_2
	\end{array} \right) +\left( \begin{array}{cc}
	f_{11} & f_{12} \\ f_{21} & -f_{11}
	\end{array} \right) \left( \begin{array}{c}
	y_1 \\ y_2
	\end{array} \right)
	=0\,, \label{eqsystemxxxyyy}
	\\ f_{11}=\dfrac{k^2 \Pi_0'}{A }+\dfrac{m(2T+m\Pi_0')}{A\vvarpi^2} +\dfrac{\rho_0^2\omega_0^2(F^2\vvarpi g-\W^2)}{AS\vvarpi }\,,
	\quad  f_{12} =\dfrac{\ktilde^2\vvarpi}{A} \,, 
	\\ f_{21}= -\dfrac{A}{\vvarpi}+\dfrac{k^2 \Pi_0'^2 }{A\vvarpi }+\dfrac{(2T+m\Pi_0')^2}{A\vvarpi^3} 
	+\rho_0\left(\dfrac{g }{\vvarpi}\right)'-\dfrac{ \rho_0^2 
	(F^2\vvarpi g-\W^2)^2}{AS\vvarpi^3} \,, 
	\nonumber \\
	\F =\bm k_0\cdot\bm B_0 \,, \quad T=FB_{0\phi}+\rho_0\omega_0V_{0\phi} \,, \quad  
	\W=\omega_0B_{0\phi}+ FV_{0\phi}\,, 
	\\
	A=\rho_0\omega_0^2-\F^2 \,, \quad
	S=\rho_0(Ac_s^2+\omega_0^2B_0^2) \,, \quad
	\ktilde^2=\dfrac{\rho_0^2\omega_0^4 }{S}-k^2-\dfrac{m^2}{\vvarpi^2}  \,. 
\end{eqnarray}
The connection of the sound velocity (only the unperturbed value is needed) with the unperturbed pressure essentially enters in the expression of $\Pi_0$.
For an ideal equation of state with internal energy density $\dfrac{P_0}{\Gamma-1}$, the sound velocity is $c_s=\sqrt{\dfrac{\Gamma P_0}{\rho_0}}$.

To simplify the expressions we define the quantities $\F$, $A$, $S$, $ \ktilde$, $T$ and $\W$. The first four also appear in the planar case and have the same meaning, see Ref.~\cite{2025Symm...17..150V}. 
Namely, $F$ is connected to magnetic tension, $S$ to compressibility, and $\ktilde$ to the local wavenumber in the $\hat\vvarpi$ direction.
(Defining $\lambda^2=\dfrac{\rho_0^2\omega_0^4 }{S}-k^2$, one can replace a planar wave $e^{i(\ktilde x+k_{0y}y+k_{0z}z-\omega t)}$ with $H^{(1)}_m(\lambda\vvarpi)e^{i(m\phi+kz-\omega t)}$ in cylindrical geometry, involving Hankel functions. The relation $\ktilde^2=\lambda^2-m^2/\vvarpi^2$ shows that $\ktilde$ splits into $\lambda$, representing the wavenumber, and $m$, the order of the Hankel function. This replacement is exact in the homogeneous case, as shown below, and gives the essential meaning of $\tilde\kappa$ and $\lambda$ in all cases.)

The Alfv\'en velocity also enters naturally in the expressions and has an important role. It is defined as $\bm v_{\rm A}=\dfrac{\bm B_0}{\sqrt{\rho_0}}$ and its component along $\bm k_0$ is $\bm v_{\rm A \parallel}=\dfrac{F/k_0}{\sqrt{\rho_0}}$.
Using this velocity we can write 
$A=\rho_0(\omega_0^2-k_0^2v_{\rm A \parallel}^2)$ and 
$S=\rho_0^2\left[(c_s^2+v_{\rm A}^2)\omega_0^2-k_0^2c_s^2v_{\rm A \parallel}^2 \right]$.

\subsection{The Principal Equation}\label{secprincipal}

As explained in Ref.~\cite{2024Univ...10..183V}, in order to find the dispersion relation of the wave or instability it is sufficient to work with the ratio $Y=\xxx/\yyy$. The system~(\ref{eqsystemxxxyyy}) implies that this ratio satisfies the principal equation 
\begin{eqnarray}
	Y' =f_{21} Y^2-2f_{11}Y-f_{12}  
	\,, \label{eqprincipal} \end{eqnarray} 
and we only need to integrate this single first-order differential equation, requiring $Y$ to be everywhere continuous and to satisfy the correct boundary conditions at the extreme values of $\vvarpi$.

Knowing $Y$ we can find all the other perturbations from 
\begin{eqnarray} 
	\dfrac{\yyy'}{\yyy}=-f_{21}Y+f_{11} \,, \quad \xxx=Y\yyy \,,
	\label{eqsforxxxyyyY}
\end{eqnarray}
together with the relations~(\ref{v1varpi6})--(\ref{pressure6}) given in Appendix~\ref{appendixA}. We emphasize, however, that these are not needed to obtain the dispersion relation.

Although the principal equation~(\ref{eqprincipal}) is the only one required, it is sometimes useful -- especially for analytical work -- to consider equivalent expressions given in Appendix~\ref{appendix-sl-oscil}. These include alternative expressions of the principal equation, as well as reformulations of the system (\ref{eqsystemxxxyyy}) as a single second-order differential equation for either $\xxx$ or $\yyy$. 
Among the many possibilities, one can write them as a Sturm-Liouville problem or as a “variable frequency oscillator” equation. 

Note that in cases with real oscillating eigenfunctions, $Y$ is real and becomes infinite at some points; the same holds for $1/Y$. This difficulty can be handled numerically by working with an angular variable, as described in Section 2 of Ref.~\cite{2024Symm...16..648V}. 
In particular, with the substitution 
$\dfrac{1}{Y}=-{\cal Y}_3\cot\dfrac{\Phi}{2}+{\cal Y}_4$, 
where ${\cal Y}_{3,4}(\vvarpi)$ are chosen functions, the infinities of $1/Y$ (i.e., the zeros of $Y$) correspond to $\Phi=2n\pi$ with integer $n$.
The principal equation then becomes a differential equation for $\Phi(\vvarpi)$ 
\begin{adjustwidth}{-\extralength}{0cm}\begin{eqnarray}
	\Phi' =2 {\cal Y}_3 f_{12} +\left(\dfrac{{\cal Y}_3'}{{\cal Y}_3}-2 {\cal Y}_4 f_{12}-2f_{11} \right) \sin\Phi  + \dfrac{({\cal Y}_4^2-{\cal Y}_3^2)f_{12} + 2{\cal Y}_4 f_{11}-f_{21}-{\cal Y}_4'}{{\cal Y}_3} (1-\cos\Phi) \,. 
	\label{eqprincphi}
\end{eqnarray}\end{adjustwidth}
In fact working with $\Phi$ instead of $Y$ may be useful for numbering the eigenvalues and eigenfunctions, as will be shown in specific examples below. 
The choice of ${\cal Y}_{3,4}(\vvarpi)$ that eliminates $\Phi$ from the right-hand side of the transformed equation is of particular interest; it corresponds to the Schwarzian $\Phi$ approach presented in Ref.~\cite{2024Symm...16..648V}.
In that case, $\Phi$ is given by the system 
\begin{eqnarray}
\Phi' =2 f_{12} {\cal Y}_3 \,, \quad  
{\cal Y}_3'=2 f_{12} {\cal Y}_3{\cal Y}_4+2f_{11}{\cal Y}_3 \,, \quad 
{\cal Y}_4'=f_{12}({\cal Y}_4^2-{\cal Y}_3^2) + 2f_{11}{\cal Y}_4 -f_{21} \,,
\label{eqsystemschwrzian}
\end{eqnarray}
(or equivalently by the Schwarzian $\left\{\cot\dfrac{\Phi}{2},\vvarpi\right\}=2\kappa_1^2$), and the solution is 
\begin{eqnarray}
\xxx\propto \dfrac{1}{\sqrt{{\cal Y}_3}}\sin\dfrac{\Phi}{2} \,, \quad 
\yyy=\dfrac{\xxx}{Y} \,, \quad 
\dfrac{1}{Y}=-{\cal Y}_3\cot\dfrac{\Phi}{2}+{\cal Y}_4 \,.
\end{eqnarray}

\subsection{Boundary Conditions}\label{secboundary}

The functions $\xxx$ and $\yyy$ are everywhere continuous, and the same holds for their ratio $Y$. Thus, in cases where the unperturbed state has tangential discontinuities, we simply continue the integration of the principal equation when passing from one medium to the next, keeping $Y$ continuous.  
\\ 
If we follow the Schwarzian approach, we integrate the system (\ref{eqsystemschwrzian}) while keeping $\Phi$, ${\cal Y}_3$, and ${\cal Y}_4$ continuous at the interfaces, which guarantees the continuity of $Y$ and $\xxx$.

When the fluid ends at some extreme value of $\vvarpi $, the value of $Y$ is known there and must be used as a boundary condition. For example, if the plasma meets a solid wall, $Y$ vanishes because the Lagrangian displacement $\xxx$ vanishes. If the plasma is in contact with a medium of constant pressure, $Y$ becomes infinity at the interface because $\yyy$ vanishes. 

Two nontrivial boundary conditions that require closer examination correspond to the symmetry axis $\vvarpi\to 0$ and to large radial distances $\vvarpi\to+\infty$. Note that the non-diverging solution of the principal equation is automatically selected when the problem is integrated numerically using the Schwarzian approach, which is a significant advantage of this method. Here, however, we obtain analytical expressions for the boundary conditions for $Y$ by simplifying the equations.

\subsubsection{Boundary Conditions on the Symmetry Axis}  

Near the axis, the azimuthal components of the velocity and the magnetic field scale linearly with the cylindrical distance ($V_{0\phi}\propto \vvarpi$, $B_{0\phi}\propto \vvarpi$).
The same holds for $T$, $\W$, and $\Pi_0'$.
We are interested in examining flows near the axis in the absence of gravity, although the following analysis also applies provided that the gravitational acceleration $ g $ vanishes at least linearly with $\vvarpi$ near the axis. 
To determine the behavior of $f_{ij}$ and of the solution near the axis, we must distinguish between the cases $m\neq 0$ and $m=0$.

For $ m\neq 0$ we get $f_{ij}=\dfrac{d_{ij}}{\vvarpi}$ with constant
$d_{11}=\displaystyle\lim_{\vvarpi \to 0} \dfrac{2mT+m^2\Pi_0'}{A\vvarpi} $,
$d_{12}=\displaystyle\lim_{\vvarpi \to 0} \dfrac{-m^2}{A} $,
$d_{21}=\displaystyle\lim_{\vvarpi \to 0} \left[-A
+\dfrac{(2T+m\Pi_0')^2 }{A\vvarpi^2}\right] $.
Changing variable to $\ln\vvarpi$, we conclude that the system~(\ref{eqsystemxxxyyy}) has power-law solutions $y_{1,2}\propto \vvarpi^{\pm\sqrt{d_{11}^2+d_{12}d_{21}}}=\vvarpi^{\pm|m|}$ with 
$\dfrac{\yyy}{\xxx}=\dfrac{d_{11}\pm|m|}{-d_{12}}$.
The physically acceptable solution is the non-diverging (upper sign), for which 
\begin{eqnarray} 
 \left.\dfrac{1}{Y}\right|_{\vvarpi=0}
=\displaystyle\lim_{\vvarpi \to 0}\dfrac{(|m|-1)A+\rho_0(\omega-kV_{0z})^2-k^2B_{0z}^2}{m^2} \mbox{ for } m\neq 0\,.
\label{boundaryaxismneq0}
\end{eqnarray}

For $m=0$ near the axis we obtain $f_{11}=\vvarpi b_{11}$, $f_{12}=\vvarpi b_{12}$, $f_{21}=\dfrac{b_{21}}{\vvarpi}$  with constant  
$b_{11}= \dfrac{k^2\Pi_0'}{A\vvarpi}+\dfrac{\rho_0^2\omega_0^2
(F^2\vvarpi g-\W^2)}{SA\vvarpi^2} $, $ b_{12} =\dfrac{\ktilde^2 }{A}$, $ b_{21}= -A
+\dfrac{4T^2 }{A\vvarpi^2} $.
The system~(\ref{eqsystemxxxyyy}) then simplifies to
$\dfrac{d(\xxx/\vvarpi^2)}{d\ln\vvarpi}+(2+b_{11}\vvarpi^2)(\xxx/\vvarpi^2)+b_{12}\yyy=0$,
$\dfrac{d\yyy}{d\ln\vvarpi}+b_{21}\vvarpi^2(\xxx/\vvarpi^2)-b_{11}\vvarpi^2\yyy=0$,
and its solution is $\yyy=$ constant, $\xxx/\vvarpi^2=-\dfrac{b_{12}}{2}\yyy$.
Thus the ratio $Y=\xxx/\yyy$ vanishes near the axis as $Y\approx -\vvarpi^2\displaystyle\lim_{\vvarpi \to 0}\dfrac{\ktilde^2}{2A}$, or, equivalently,
\begin{eqnarray} 
	Y|_{\vvarpi\approx 0}\approx -\vvarpi^2\displaystyle\lim_{\vvarpi \to 0} \dfrac{1}{2\rho_0}\dfrac{\omega_0^2-k^2c_s^2}{(v_{\rm A}^2+c_s^2)\omega_0^2-k^2c_s^2v_{\rm A}^2}  \mbox{ for } m= 0\,.
	\label{boundaryaxismeq0}
\end{eqnarray}
For practical purposes, it is sufficient to impose the boundary condition $Y|_{\vvarpi=0}=0$ for $m=0$.


\subsubsection{Boundary Conditions at Infinity}  

In the limit $\vvarpi\to+\infty$, the perturbations of the physical quantities must vanish. This condition is fulfilled provided that the radial Lagrangian displacement $y_1/\vvarpi$ and the perturbation of the total pressure $y_2$ decay asymptotically. Furthermore, when $y_{1,2}$ correspond to wave-like solutions, they must represent outgoing waves in this limit.

An interesting subcase corresponds to homogeneous static plasma with zero $B_{0\phi}$ (and no gravity). 
In that case    
$f_{11}=0$, $f_{12} =\dfrac{\lambda^2\vvarpi}{A}-\dfrac{m^2}{A\vvarpi}$, $ f_{21}= -\dfrac{A}{\vvarpi} $, with constant $A=\rho_0(\omega^2-k^2v_{\rm A}^2)$, and constant $\lambda^2
=\dfrac{(\omega^2-k^2v_{\rm A}^2)(\omega^2-k^2c_s^2)}{(v_{\rm A}^2+c_s^2)\omega^2-k^2c_s^2v_{\rm A}^2}$.
(The quantity $\lambda$ and its connection to $\ktilde$ through $\lambda^2=\ktilde^2+m^2/\vvarpi^2$ has already been discussed.)
The system~(\ref{eqsystemxxxyyy}) then simplifies to
$\vvarpi\xxx'+(\lambda^2\vvarpi^2-m^2)\yyy/A=0$, $\vvarpi\yyy'/A-\xxx=0$.
Combining these, we find that $\yyy$ satisfies the Bessel equation 
$\vvarpi^2\yyy''+\vvarpi\yyy'+(\lambda^2\vvarpi^2-m^2)\yyy=0$, while 
$Y=\dfrac{\vvarpi\yyy'}{A\yyy}$.
Using standard properties of the Bessel functions, we choose for $\yyy$ the Hankel function whose amplitude decreases with distance.
Selecting the root of $\lambda^2 $ with positive imaginary part, 
$\lambda =i\sqrt{\dfrac{(k^2v_{\rm A}^2-\omega^2)(k^2c_s^2-\omega^2)}{k^2c_s^2v_{\rm A}^2-(v_{\rm A}^2+c_s^2)\omega^2}}$ (taking the principal value of the root), the solution is 
$\yyy\propto H_{|m|}^{(1)}(\lambda\vvarpi)$, which represents an outgoing wave if $\Re\lambda$ and $\Re\omega$ have the same sign. 
The resulting solution of the principal equation is 
\begin{eqnarray} 
	Y=\dfrac{\lambda\vvarpi\dfrac{ H_{|m|+1}^{(1)}(\lambda\vvarpi)}{H_{|m|}^{(1)}(\lambda\vvarpi)}-|m|}{\rho_0(k^2v_{\rm A}^2-\omega^2)} 
	\,,
	\quad \lambda =i\sqrt{\dfrac{(k^2v_{\rm A}^2-\omega^2)(k^2c_s^2-\omega^2)}{k^2c_s^2v_{\rm A}^2-(v_{\rm A}^2+c_s^2)\omega^2}} \,.
	\label{boundaryinftybessel}
\end{eqnarray}
An alternative expression, more convenient when $\lambda$ is purely imaginary
(using the relation between Hankel and modified Bessel functions $H_\alpha^{(1)}(\lambda\vvarpi)=\dfrac{2}{i^{\alpha+1}\pi}K_\alpha(-i\lambda\vvarpi)$) is 
\begin{eqnarray} 
	Y=\dfrac{-i\lambda\vvarpi\dfrac{ K_{|m|+1}(-i\lambda\vvarpi)}{K_{|m|}(-i\lambda\vvarpi)}-|m|}{\rho_0(k^2v_{\rm A}^2-\omega^2)} 
	\,,
	\quad -i\lambda =\sqrt{\dfrac{(k^2v_{\rm A}^2-\omega^2)(k^2c_s^2-\omega^2)}{k^2c_s^2v_{\rm A}^2-(v_{\rm A}^2+c_s^2)\omega^2}} \,.
	\label{boundaryinftybesselK}
\end{eqnarray}

\subsubsection{Summary of Equations and Boundary Conditions}  
The equations and boundary conditions required to apply the minimalist approach and obtain the dispersion relation for a given problem in cylindrical geometry are summarized in Table~\ref{tab:formulas}. 

\begin{table}[h!] \centering \begin{tabular}{m{1.72cm} m{11.28cm}} 
	\hline Principal equation: & \( Y' =f_{21} Y^2-2f_{11}Y-f_{12} \)  
\\ with & \smallskip \( f_{11}=\dfrac{k^2 \Pi_0'}{A }+\dfrac{m(m\Pi_0'+2T)}{\vvarpi^2 A} +\dfrac{\omega_0^2(F^2\vvarpi g-\W^2)}{\vvarpi AS/\rho_0^2} \,, \quad f_{12} =\dfrac{\ktilde^2\vvarpi}{A} \,,\) 
\\ &  \( 
f_{21}=-\dfrac{A}{\vvarpi}+\dfrac{k^2 \Pi_0'^2 }{\vvarpi A}+\dfrac{(m\Pi_0'+2T)^2}{\vvarpi^3 A} 
+\rho_0\left(\dfrac{ g }{\vvarpi}\right)'-\dfrac{  (F^2\vvarpi g-\W^2)^2}{\vvarpi^3 AS/\rho_0^2}\,, \)
\\ \multicolumn{2}{c}{ \( 
	\omega_0=\omega-kV_{0z}-\dfrac{m}{\vvarpi}V_{0\phi} \,, \  \F =kB_{0z}+\dfrac{m}{\vvarpi}B_{0\phi}  \,, \ T=FB_{0\phi}+\rho_0\omega_0V_{0\phi} \,, \
	\W=\omega_0B_{0\phi}+ FV_{0\phi}\,, \) }
\\ 
\multicolumn{2}{c}{ \( 
	A=\rho_0\left(\omega_0^2-{\F^2}/{\rho_0}\right) \,, \
	{S}/{\rho_0^2}= \left(c_s^2+{B_0^2}/{\rho_0}\right)\omega_0^2-c_s^2{\F^2}/{\rho_0}  \,, \
	\ktilde^2=\dfrac{\omega_0^4 }{S/\rho_0^2}-k^2-\dfrac{m^2}{\vvarpi^2} \,,
	\) } 
\\ 
\multicolumn{2}{l}{  \( 
	\phantom{\left(\dfrac{AAAA11}{B}\right)}
 \Pi_0'
	= -\rho_0 g_{\rm eff}-\dfrac{B_{0\phi}^2}{\vvarpi} 
\,, \ \Pi_0=\dfrac{\rho_0 c_s^2}{\Gamma}+\dfrac{B_0^2}{2}\,,  \ g_{\rm eff}=g-\dfrac{V_{0\phi}^2}{\vvarpi}\,. 
	\) }
\\ \hline 
\smallskip Boundary conditions: & \( Y  \) continuous everywhere,
	\\ \cline{2-2} & \smallskip 
	if $ m\neq 0$ on the axis  \( Y|_{\vvarpi=0} 
	=\displaystyle\lim_{\vvarpi \to 0}\dfrac{m^2}{(|m|-1)A+\rho_0(\omega-kV_{0z})^2-k^2B_{0z}^2} \),
	\\ \cline{2-2} & \smallskip 
	if $ m= 0$ near the axis $ Y|_{\vvarpi\approx 0}\approx -\vvarpi^2\displaystyle\lim_{\vvarpi \to 0} \dfrac{\ktilde^2}{2A} $\,,
	\\ \cline{2-2} & \smallskip 
	in the regime that includes $\vvarpi=\infty$,
	assuming static, homogeneous plasma
	\\   & \smallskip  with $B_{0\phi}=0$\,, \quad 
	$Y=\dfrac{1}{A}\left[|m|-\lambda\vvarpi\dfrac{ H_{|m|+1}^{(1)}(\lambda\vvarpi)}{H_{|m|}^{(1)}(\lambda\vvarpi)}\right] , \quad \lambda =i\sqrt{k^2-\dfrac{\rho_0^2\omega^4}{S}}$\,.
	\\ \hline \end{tabular}
\caption{Minimalist approach equations and boundary conditions in cylindrical geometry} \label{tab:formulas} \end{table}

\section{Wall-Bounded Rotational Instabilities}\label{secwall-b}

Wall-bounded instabilities refer to configurations in which the flow is confined between two solid cylindrical surfaces (with the inner boundary possibly coinciding with the symmetry axis and the outer boundary extending up to a finite radius or even to $\vvarpi\to\infty$). As an illustrative example, consider a magnetized incompressible fluid rotating between two cylindrical walls $\vvarpi_1<\vvarpi<\vvarpi_2$, with angular velocity $\Omega=\dfrac{V_{0\phi}}{\vvarpi}$ and an axial magnetic field $B_0\hat z$.
The radial profile of $\Omega$ is determined by the gravitational and pressure fields, and force balance in the unperturbed state requires $\Pi_0'=-\rho_0 g_{\rm eff}$, where $g_{\rm eff}=g-\Omega^2\vvarpi$ incorporates centrifugal acceleration.  
We focus here on axisymmetric perturbations ($m=0$) in the incompressible limit ($S\to\infty$), for which we find \( f_{11}=-\dfrac{k^2\rho_0 g_{\rm eff}}{A } \),  \(f_{12} =-\dfrac{k^2\vvarpi}{A} \),  \( f_{21}=-\dfrac{A}{\vvarpi}+\dfrac{k^2 \rho_0^2 g_{\rm eff}^2 }{A\vvarpi }+\dfrac{4\rho_0^2\omega^2\Omega^2}{A\vvarpi} +\rho_0\left(\dfrac{g}{\vvarpi}\right)' \),   
\( A=\rho_0 (\omega^2-k^2v_{\rm A}^2) \).

\subsection{Approximate Dispersion Relation Using WKBJ}\label{secappinc}
Before presenting exact results, we briefly discuss the approximate dispersion relation obtained from the WKBJ approximation. This provides expressions that capture the correct qualitative behavior of the instabilities and aid physical interpretation. The approximation is based on partially neglecting spatial variations of the unperturbed state, while retaining variations in the quantities that control the relevant physical mechanisms. Density variations may be unimportant, but they must be retained in terms involving the effective gravity if buoyancy effects are to be captured—this is the essence of the Boussinesq approximation. Likewise, variations of the specific angular momentum $\vvarpi^2\Omega$ must be kept when studying CFI and MRI, since these mechanisms depend sensitively on the radial gradient of angular momentum.

Based on the ``variable frequency'' equation~(\ref{odefory1oscillator}), the local radial wavenumber is $ k_\vvarpi=\kappa_1$, with $\kappa_1$ given by equation~(\ref{kappa1oscillator}).  
Ignoring spatial variations of $v_{\rm A}$ we obtain 
$k_\vvarpi^2
=-k^2-\dfrac{3}{4\vvarpi^2}+\dfrac{\rho_0'}{2\vvarpi\rho_0}+\dfrac{\rho_0'^2}{4\rho_0^2}-\dfrac{\rho_0''}{2\rho_0}-\dfrac{ \dfrac{(\vvarpi^4 \Omega^2)'}{\vvarpi^3} -\dfrac{\rho_0'g_{\rm eff}}{\rho_0} }{v_{\rm A}^2-\omega^2/k^2} +\dfrac{4\Omega^2v_{\rm A}^2}{(v_{\rm A}^2-\omega^2/k^2)^2 }$.
Following the Boussinesq approximation, and assuming $|k|\vvarpi\gg 1$ we obtain the approximate expression  
$ k_\vvarpi^2
=-k^2 -\dfrac{ \dfrac{(\vvarpi^4 \Omega^2)'}{\vvarpi^3} -\dfrac{\rho_0'g_{\rm eff}}{\rho_0} }{v_{\rm A}^2-\omega^2/k^2} +\dfrac{4\Omega^2v_{\rm A}^2}{(v_{\rm A}^2-\omega^2/k^2)^2 }
$, giving the local $k_\vvarpi$ as a function of $\vvarpi$. In cases where the radial wavenumber does not change significantly over the interval $\vvarpi_1<\vvarpi<\vvarpi_2$ -- that is, when $\kappa_1'/\kappa_1^2$ is small and $k_\vvarpi$ may be treated as the constant wavenumber of standing waves in the radial direction, equal to an integer multiple of $\dfrac{\pi}{\vvarpi_2-\vvarpi_1}$ -- the above relation can be interpreted as an approximate dispersion relation. Under these assumptions it can be written in the form of the following quadratic 
\begin{eqnarray} 
\dfrac{k_\vvarpi^2+k^2}{k^2}(\omega^2-k^2v_{\rm A}^2)^2
- \left[\dfrac{(\vvarpi^4\Omega^2)'}{\vvarpi^3} -\dfrac{\rho_0'g_{\rm eff}}{\rho_0}\right](\omega^2-k^2v_{\rm A}^2)-4\Omega^2k^2v_{\rm A}^2=0 \,.
\label{eqmri} \end{eqnarray}

In the absence of magnetic fields, for constant density, and in the limit of large wavenumber ($|k|\gg |k_\vvarpi|$), we recover the familiar result of radial oscillations with the epicyclic frequency $\omega=\sqrt{\dfrac{(\vvarpi^4\Omega^2)'}{\vvarpi^3}}$, or CFI with growth rate $\Im\omega=\sqrt{-\dfrac{(\vvarpi^4\Omega^2)'}{\vvarpi^3}}$, depending on the sign of $(\vvarpi^4\Omega^2)'$, i.e., the distribution of angular momentum in the flow.
Similarly, in the unmagnetized and non-rotating case we recover the standard result of radial oscillations with the buoyancy frequency $\omega=\sqrt{-\dfrac{\rho_0'g_{\rm eff}}{\rho_0}}$, or instability when the stratification is unstable.
With a nonzero magnetic field and constant density, the necessary condition for instability follows immediately from the approximate relation: $(\Omega^2)'<0$, and unstable modes occur for wavenumbers 
$k_\vvarpi^2+k^2<\dfrac{-\vvarpi(\Omega^2)'}{v_{\rm A}^2}$.

A thin astrophysical disk around a central mass $M$ can be approximated as a cylindrical flow in a gravitational field $g=\dfrac{GM}{\vvarpi^2}$ with Keplerian rotation $\Omega=\sqrt{\dfrac{GM}{\vvarpi^3}}$.
Hydrodynamic perturbations in such a disk are stable, with oscillation frequency $\omega=\Omega$. 
In the presence of magnetic field, the solutions of the dispersion relation~(\ref{eqmri}) take the form $\omega^2=k^2v_{\rm A}^2+\dfrac{\Omega^2}{2}\pm\sqrt{\dfrac{\Omega^4}{4}+4k^2v_{\rm A}^2\Omega^2}$ and include an unstable branch ($\omega^2<0$) provided that $k^2<\dfrac{3\Omega^2}{v_{\rm A}^2}$.
This is the MRI, first analyzed for magnetized Couette flow in Ref.~\cite{1960PNAS...46..253C}  and later identified as the key instability in magnetized astrophysical disks in Ref.~\cite{1991ApJ...376..214B}.

For a general $\Omega(\vvarpi)$ profile, the Rayleigh criterion $(\vvarpi^4\Omega^2)'<0$, meaning that the specific angular momentum decreases outward, is associated with CFI, which appears already in the hydrodynamic case. By contrast, the condition $(\Omega^2)'<0$, even when the specific angular momentum increases, is associated with MRI, which requires the presence of a magnetic field; see the related discussion in Section 6.2 of Ref.~\cite{2019MNRAS.488.4061K}. 
Buoyancy instability may also occur. In the incompressible limit it is linked to unstable density stratification with respect to the effective gravity, i.e. when $-\dfrac{\rho_0'g_{\rm eff}}{\rho_0}<0$.

In the general incompressible case, only the smaller root of the quadratic dispersion relation~(\ref{eqmri}) can become negative, i.e. the
\begin{adjustwidth}{-\extralength}{0cm}\begin{eqnarray} 
	\omega^2=k^2v_{\rm A}^2+\dfrac{(\vvarpi^4\Omega^2)'/\vvarpi^3-\rho_0'g_{\rm eff}/\rho_0}{2(1+k_\vvarpi^2/k^2)}
	-\sqrt{\left[\dfrac{(\vvarpi^4\Omega^2)'/\vvarpi^3-\rho_0'g_{\rm eff}/\rho_0}{2(1+k_\vvarpi^2/k^2)}\right]^2+\dfrac{4\Omega^2 k^2v_{\rm A}^2}{1+k_\vvarpi^2/k^2}} \,.
	\label{eqmriincompr1}
\end{eqnarray}\end{adjustwidth}
It is straightforward to show that, for nonzero magnetic field, this root becomes negative for wavenumbers satisfying
$ k_\vvarpi^2+k^2<\dfrac{\rho_0'g_{\rm eff}/\rho_0-\vvarpi(\Omega^2)'}{v_{\rm A}^2}$,
provided that the necessary condition for instability 
$-\rho_0'g_{\rm eff}/\rho_0+\vvarpi(\Omega^2)'<0
\Leftrightarrow (\rho_0\Omega^2)'<\rho_0'g/\vvarpi$ is satisfied. 
For zero magnetic field, the result reduces to 
\begin{eqnarray} 
\omega^2=\dfrac{(\vvarpi^4\Omega^2)'/\vvarpi^3-\rho_0'g_{\rm eff}/\rho_0}{2(1+k_\vvarpi^2/k^2)}
-\left|\dfrac{(\vvarpi^4\Omega^2)'/\vvarpi^3-\rho_0'g_{\rm eff}/\rho_0}{2(1+k_\vvarpi^2/k^2)}\right| \,, 
\nonumber \end{eqnarray}
corresponding to instability for all wavenumbers if $\dfrac{(\vvarpi^4\Omega^2)'}{\vvarpi^3}-\dfrac{\rho_0'g_{\rm eff}}{\rho_0}<0$.)

\subsection{Exact Dispersion Relation Using the Principal Equation}\label{sectwocylindinc}
The previous results were based on a local analysis. 
The exact dispersion relation in a rotating flow whose properties vary with radius is, however, a global eigenvalue problem.  

To obtain the exact dispersion relation, we must first to specify the unperturbed state in detail.
As an illustrative example, consider an incompressible fluid confined between two cylindrical walls $\vvarpi_1<\vvarpi<\vvarpi_2$, rotating with power-law angular velocity in the absence of gravity and possessing uniform density and magnetic field. 
We normalize all quantities using $\vvarpi_2$ as the unit length, $\Omega|_{\vvarpi_2}$ as the unit frequency, and the fluid density as the unit density. Under these normalizations the flow occupies the interval $1-\Delta\vvarpi<\vvarpi<1$, has density $\rho=1$, magnetic field $v_{\rm A}\hat z$, and rotates with $\Omega=\vvarpi^b$.
The three parameters $\Delta\vvarpi$, $b$ and $v_{\rm A}$ uniquely define the unperturbed configuration.

To determine the growth rates of the instability, we integrate the principal equation with boundary conditions $Y|_{\vvarpi_{1,2}}=0$
on the cylindrical walls.
We restrict attention to axisymmetric perturbations, for which the eigenvalues are real ($\omega^2=-\Im\omega^2$, negative for instability) and the eigenfunctions are oscillatory. To avoid singular behavior at points where $Y$ vanishes, we employ the Schwarzian approach (see the end of Section~\ref{secprincipal}).
Starting the integration from the inner cylinder $\vvarpi_1$ with $\Phi|_{\vvarpi_1}=0$, the eigenvalues are those for which the solution satisfies
$\Phi|_{\vvarpi_2}=2n\pi$, with integer $n$, upon reaching the outer cylinder. 

\begin{figure}
	\begin{minipage}[l]{.45\textwidth}
		\includegraphics[width=1\textwidth]{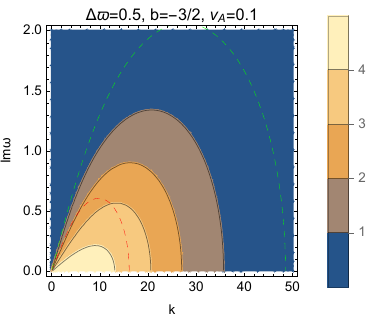}
	\end{minipage}
	\hfill
	 \begin{minipage}[l]{.55\textwidth}
	 	\includegraphics[width=1\textwidth]{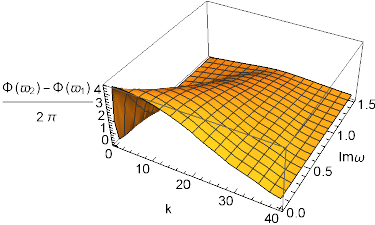}
	 \end{minipage}
\\
	\includegraphics[width=.62\textwidth]{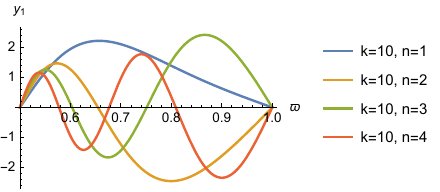}
	\hfill
	\includegraphics[width=.38\textwidth]{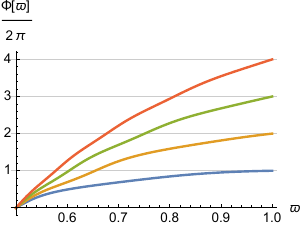}  
	\caption{The instability in a wall-bounded rotating flow.
	The top row shows $\dfrac{\Phi|_{\vvarpi_2}-\Phi|_{\vvarpi_1}}{2\pi}$ as a function of $k$--$\Im\omega$,
	together with its isocontours, which form the branches of the dispersion relation. Thin dashed curves correspond to the approximate dispersion relation~(\ref{eqmri}) with $k_\vvarpi=\pi/\Delta\vvarpi$; the green curve uses values at $\vvarpi_1$, while the red curve uses values at $\vvarpi_2$. 
	The bottom row shows $\xxx$ (with arbitrary normalization) and $\Phi$ for $k=10$ and the four eigenvalues $\Im\omega\approx 1.02, 0.76, 0.52, 0.21$.
	All quantities are normalized using $\vvarpi_2$ as the unit length and $\Omega|_{\vvarpi_2}$ as the unit frequency.
	\label{figva01b1.5r05}
	}

\vspace{5mm}	
\includegraphics[width=.45\textwidth]{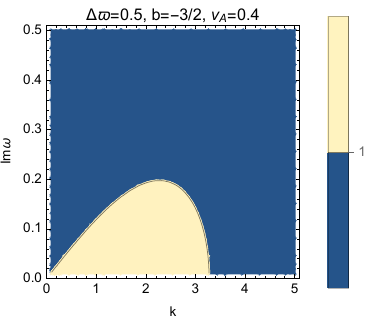}
\hfill
\includegraphics[width=.45\textwidth]{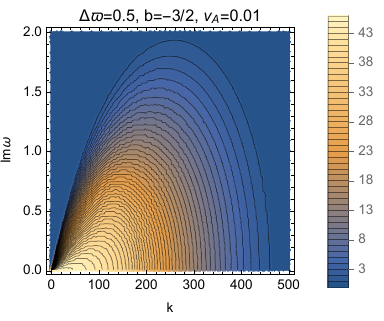} 
\\ 
\includegraphics[width=.45\textwidth]{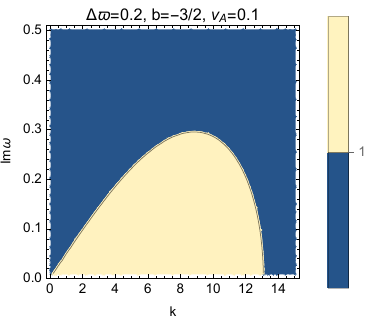}
\hfill
\includegraphics[width=.45\textwidth]{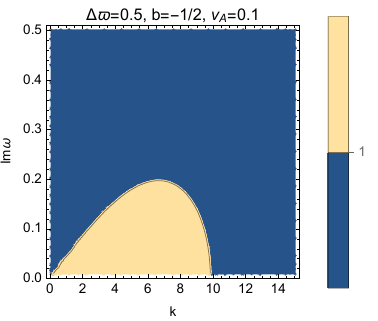}  
\caption{The instability in a wall-bounded rotating flow for other parameter combinations.
	\label{figva01b1.5r05others}
}
\end{figure}

By performing the integration over a grid in the ($k$--$\Im\omega$) plane, we obtain the quantity
$\dfrac{\Phi|_{\vvarpi_2}-\Phi|_{\vvarpi_1}}{2\pi}$ as function of $k$ and $\Im\omega$.
The contours where this function equals an integer $n$ form the branches of the dispersion relation. 
An example is shown in Figure~\ref{figva01b1.5r05} for $\Delta\vvarpi=0.5$, $b=-3/2$, and $v_{\rm A}=0.1$. As seen in the top-left panel, four branches exist in this case, corresponding to $n=1,2,3,4$, as indicated by the color bar.
Because
$\Phi(\vvarpi)$ is monotonically increasing function, the integer $n$ can be used to number the eigenvalues/eigenfunctions. 
Indeed, as shown in the bottom-left panel of Figure~\ref{figva01b1.5r05}, the number of zeros of $\xxx$ (and $Y$) in the interior of the flow equals $n-1$. 

The magnetic field, through the stabilizing effect of its tension, strongly modifies both the growth rate of the instability and the range of wavenumbers over which it occurs. The entire surface $\dfrac{\Phi|_{\vvarpi_2}-\Phi|_{\vvarpi_1}}{2\pi}$ is shifted toward lower $\Im\omega$ and lower $k$, as seen for example in the top-left panel of Figure~\ref{figva01b1.5r05others}. A further increase to values $v_{\rm A}\gtrsim 0.45$ (with the other parameters unchanged) completely stabilizes the flow. 
Conversely, decreasing $v_{\rm A}$ increases the growth rate and, more importantly, enlarges both the number of branches and the range of unstable wavenumbers, as shown in the top‑right panel of Figure~\ref{figva01b1.5r05others}.

The growth rate also decreases as the width $\Delta\vvarpi$ becomes smaller, because the variation of $\Omega$ across the flow -- which drives the instability -- becomes weaker. An example is shown in the bottom-left panel of Figure~\ref{figva01b1.5r05others}. A further reduction to values $\Delta\vvarpi\lesssim 0.16$ (with the other parameters unchanged) completely stabilizes the flow. 

The variation of $\Omega$ across the flow also depends on the exponent $b$. As expected, smaller $|b|$ stabilizes, since it reduces the difference $\Omega|_{\vvarpi_1}-\Omega|_{\vvarpi_2}$. An example is shown in the bottom-right panel of Figure~\ref{figva01b1.5r05others}. Decreasing $|b|$ to values $|b|\lesssim 0.2$ (with the other parameters unchanged) again leads to complete stabilization.

All of these behaviors are characteristic of the MRI. (The chosen value of $b$ does not correspond to CFI, and the absence of density stratification excludes buoyancy modes.)

The exact results can be qualitatively understood using the WKBJ analysis of Section~\ref{secappinc}. 
The range of unstable wavenumbers and the number of branches are controlled by the condition $k_\vvarpi^2+k^2<k_{\rm max}^2$ with $k_{\rm max}^2=\dfrac{\rho_0'g_{\rm eff}/\rho_0-\vvarpi(\Omega^2)'}{v_{\rm A}^2}$, combined with the approximate standing-wave relation $k_\vvarpi=\dfrac{n\pi}{\Delta\vvarpi}$ obtained by approximating the eigenfunction as $\sin(k_\vvarpi \vvarpi+C)$ and enforcing vanishing at both boundaries.
This instability condition gives the maximum $k$ of each branch (for the first branch it is roughly the $k_{\rm max}$ above) and also the maximum mode number $n$ (i.e., the number of branches) through $n<\dfrac{k_{\rm max}\Delta\vvarpi}{\pi}$. 
These estimates are necessarily rough, since $k_{\rm max}=\dfrac{\sqrt{\rho_0'g_{\rm eff}/\rho_0-\vvarpi(\Omega^2)'}}{v_{\rm A}}$ depends on radius. In the present case, $k_{\rm max}=\dfrac{\sqrt{-2b}}{v_{\rm A}\vvarpi^{-b}}$, so it is not obvious which value should be used. One may employ mean values, or obtain upper and lower bounds using the extremal values of the functions entering the WKBJ expressions. The same applies to the dispersion relation itself.
The WKBJ result is 
$\omega^2=k^2v_{\rm A}^2+\dfrac{(4+2b)\vvarpi^{2b} }{2(1+k_\vvarpi^2/k^2)}
-\sqrt{\left[\dfrac{(4+2b)\vvarpi^{2b}}{2(1+k_\vvarpi^2/k^2)}\right]^2+\dfrac{4\vvarpi^{2b}k^2v_{\rm A}^2}{1+k_\vvarpi^2/k^2}}$ with $k_\vvarpi=\dfrac{n\pi}{\Delta\vvarpi}$. The first branch is shown with dashed lines in the top-left panel of Figure~\ref{figva01b1.5r05}, using values at the inner and outer radii. As expected, it reproduces only the qualitative behavior of the exact result, which lies between these two extreme WKBJ curves.

\begin{figure}
	\includegraphics[width=.4\textwidth]{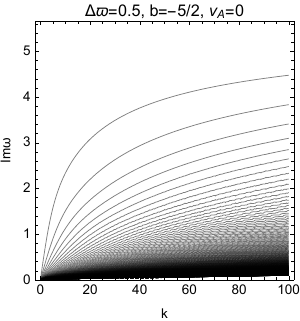}
	\hfill
	\includegraphics[width=.4\textwidth]{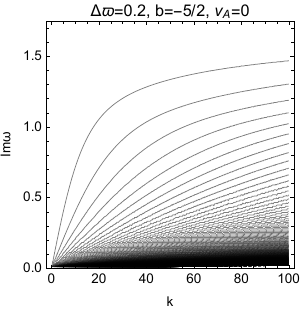}  
	\caption{The CFI in a wall-bounded rotating hydrodynamic flow.
	\label{fighydroincomr}
	}
\end{figure}  
Results for the hydrodynamic case are shown in Figure~\ref{fighydroincomr}. 
As expected from the Rayleigh criterion, CFI appears for $b<-2$. The growth rate agrees qualitatively with the WKBJ analysis of Section~\ref{secappinc}, $\omega^2=\dfrac{ {2b+4}\, }{1+k_\vvarpi^2/k^2}\Omega^2 $.
It is also evident that,  in the absence of magnetic tension, nothing limits the instability at high wavenumbers, and all CFI branches extend to $k\to\infty$. 

\section{Interface-Driven Rotational Instabilities}\label{secinterface}
A discontinuity in the unperturbed state may trigger an instability. 
Here we are interested in examining the stability of two rotating incompressible fluids that are in contact at some cylindrical radius $\rin$ in the absence of gravity. In what follows, subscript 1 refers to the inner region $\vvarpi<\rin$, which is assumed to extend to the symmetry axis, and subscript 2 to the outer $\vvarpi>\rin $, which is assumed to extend to infinity.
For simplicity, we assume uniform rotation with angular velocities $\Omega_1$ and $\Omega_2$ in the two regions, constant densities $\rho_1$ and $\rho_2$, and uniform magnetic fields corresponding to constant Alfv\'en velocities $v_{\rm A 1}$ and $v_{\rm A 2}$. We restrict attention to axisymmetric perturbations.
Under these assumptions, the functions entering the principal equation take the form \( f_{11}=\dfrac{k^2 \Omega^2\vvarpi}{\omega^2-k^2v_{\rm A}^2} \),  \(f_{12} =-\dfrac{k^2\vvarpi}{\rho (\omega^2-k^2v_{\rm A}^2)} \),  \( f_{21}= \rho \dfrac{k^2 \Omega^4\vvarpi^2+4\omega^2\Omega^2-(\omega^2-k^2v_{\rm A}^2)^2}{\vvarpi(\omega^2-k^2v_{\rm A}^2)} \), 
with the appropriate choice of parameters in each region. 

\subsection{Exact Dispersion Relation Using the Principal Equation}
We can obtain the dispersion relation by numerically integrating the principal equation. 
A particularly effective method is to use the Schwarzian approach and determine the eigenvalues from the requirement that $\dfrac{\Phi|_2-\Phi|_1}{2\pi}$
be an integer. The subscripts 1 and 2 refer to the extreme values of $\vvarpi$ in the entire plasma domain; in the present configuration these are $\vvarpi_1=0$ and $\vvarpi_2=+\infty$. This approach has two advantages. 
First, as in the wall-bounded case, the integer $\dfrac{\Phi|_{\vvarpi=\infty}-\Phi|_{\vvarpi=0}}{2\pi}$ can be used to number the eigenvalues.
Second, when the flow extends to asymptotic distances -- either toward the rotation axis or toward infinity, as happens in the present two-fluid configuration -- the correct boundary conditions are automatically satisfied. We start the integration at the interface ($\vvarpi=\rin$) and integrate both inward toward the axis and outward toward infinity, using arbitrary but identical initial values of $\Phi$ and ${\cal Y}_{3,4}$ at $\rin$. The dispersion relation corresponds to the condition that $\Phi|_{\vvarpi=\infty}-\Phi|_{\vvarpi=0}$ be an integer multiple of $2\pi$.
In practice, the integration need not reach exactly $\vvarpi=0$; it suffices to stop once $\Phi$ has approached its asymptotic constant value. Similarly, the outward integration is terminated at a sufficiently large radius where $\Phi$ has approached its asymptotic constant value.

\subsection{Analytical Dispersion Relation}\label{secincomprtwoflows}

The principal equation for axisymmetric perturbations of a homogeneous incompressible fluid (with constant density $\rho $ and magnetic field corresponding to constant Alfv\'en velocity $v_{\rm A}\hat z$) rotating uniformly (with constant angular velocity) is
\begin{eqnarray}  
		\dfrac{d }{d\vvarpi } \left(\dfrac{1}{Y}\right) =-\dfrac{k^2\vvarpi}{\rho (\omega^2-k^2v_{\rm A}^2)}\left(\dfrac{1}{Y}-\rho\Omega^2\right)^2
		+\dfrac{\rho(\omega^2-k^2v_{\rm A}^2)}{\vvarpi}
		-\dfrac{4 \rho\omega^2 \Omega^2/\vvarpi
		}{\omega^2-k^2v_{\rm A}^2 } 
		\,,
		\label{eqprincipalincomprinterf}
\end{eqnarray}
and can be solved analytically by employing the equivalent Sturm-Liouville formulation~(\ref{odefory1}).
With $\dfrac{1}{Y}=\dfrac{-f_{11}-\xxx'/\xxx}{f_{12}}= \rho\Omega^2 +\dfrac{ A\xxx'}{k^2\vvarpi \xxx}$ 
the equation for $\xxx$ is 
$\xxx''- \dfrac{\xxx'}{\vvarpi}-k^2\xxx+ \dfrac{4\Omega^2/\omega^2 }{(1-k^2v_{\rm A}^2/\omega^2)^2}k^2\xxx=0$, and can be transformed to a modified Bessel equation 
$\vvarpi^2\left(\dfrac{\xxx}{\vvarpi}\right)''+\vvarpi\left(\dfrac{\xxx}{\vvarpi}\right)'-\left(1+ \Lambda^2\vvarpi^2\right)\left(\dfrac{\xxx}{\vvarpi}\right)=0$, with $\Lambda=\sqrt{k^2-\dfrac{4\omega^2\Omega^2k^2}{(k^2v_{\rm A}^2-\omega^2)^2} }$
($\Lambda^2$ equals $-\kappa_0^2$, and $\kappa_0^2$ is negative in this case -- to avoid using the complex $\kappa_0$ we define $\Lambda$ which is real). 
The general solution is $\xxx=C_+\vvarpi K_1(\Lambda\vvarpi)+C_-\vvarpi I_1(\Lambda\vvarpi)$, and therefore the general solution of the principal equation is 
$\dfrac{1}{Y}=\rho \Omega^2
+\dfrac{ A \Lambda }{k^2\vvarpi }\, \dfrac{C_-  I_0(\Lambda\vvarpi)-C_+ K_0(\Lambda\vvarpi)}{C_- I_1(\Lambda\vvarpi)+C_+ K_1(\Lambda\vvarpi)}$.

We can use the previous result to obtain the dispersion relation for two uniformly rotating incompressible fluids in contact. In each region we select the modified Bessel function that is regular at the corresponding asymptotic limit: $I_1(\Lambda\vvarpi)$ for the inner region (regular at $\vvarpi\to 0$) and $K_1(\Lambda\vvarpi)$ for the outer region (regular at $\vvarpi\to\infty$). Thus the solution is
\begin{adjustwidth}{-\extralength}{0cm}\begin{eqnarray}
		\dfrac{1}{Y}=\left\{
		\begin{array}{ccc} 
			\rho_1\Omega_1^2
			-\dfrac{\rho_1(k^2v_{\rm A 1}^2-\omega^2)\Lambda_1 }{k^2\vvarpi }\dfrac{I_0(\Lambda_1\vvarpi)}{I_1(\Lambda_1\vvarpi)} 
			\,, & \vvarpi<\rin \,,
			& \Lambda_1= \sqrt{k^2-\dfrac{ 4\omega^2\Omega_1^2k^2}{(k^2v_{\rm A 1}^2-\omega^2)^2} } \,,
			\\ 
			\rho_2\Omega_2^2
			+\dfrac{\rho_2(k^2v_{\rm A 2}^2-\omega^2) \Lambda_2}{k^2\vvarpi }\dfrac{K_0(\Lambda_2\vvarpi)}{ K_1(\Lambda_2\vvarpi)} 
			\,, & \vvarpi>\rin \,,
			& \Lambda_2= \sqrt{k^2-\dfrac{4\omega^2\Omega_2^2k^2}{(k^2v_{\rm A 2}^2-\omega^2)^2} } \,.
		\end{array}
		\right. 
		\label{eqdispcentrtwo}
\end{eqnarray}\end{adjustwidth}
The dispersion relation results from the continuity of $Y$
\begin{adjustwidth}{-\extralength}{0cm}\begin{eqnarray}
\rho_1\Omega_1^2\rin-\rho_2\Omega_2^2\rin=
\rho_1\dfrac{k^2v_{\rm A 1}^2-\omega^2}{k^2} \Lambda_1\dfrac{I_0(\Lambda_1\rin)}{I_1(\Lambda_1\rin)}
+\rho_2\dfrac{k^2v_{\rm A 2}^2-\omega^2}{k^2}\Lambda_2\dfrac{K_0(\Lambda_2\rin)}{ K_1(\Lambda_2\rin)}  \,. 
\label{eqdispcentrrel}
\end{eqnarray}\end{adjustwidth}

\subsection{Approximate Dispersion Relation}

It is instructive to simplify the above in the limit of large $|k|\rin $, using the asymptotic approximations $K_0\approx K_1$ and $I_0\approx I_1$. The resulting approximate dispersion relation is
\begin{adjustwidth}{-\extralength}{0cm}\begin{eqnarray} 
		\rho_1\Omega_1^2\rin -\rho_2\Omega_2^2\rin 
		=\rho_1 \dfrac{k^2v_{\rm A 1}^2-\omega^2}{|k|} \sqrt{1-\dfrac{ 4\omega^2\Omega_1^2 }{(k^2v_{\rm A 1}^2-\omega^2)^2} } +\rho_2\dfrac{k^2v_{\rm A 2}^2-\omega^2}{|k|} \sqrt{1-\dfrac{4\omega^2\Omega_2^2 }{(k^2v_{\rm A 2}^2-\omega^2)^2} }
		\,. \label{eqdispcentrrelapprox}
\end{eqnarray}\end{adjustwidth}
It can be written as a quartic equation in $\omega$ and solved analytically (selecting a posteriori the roots that satisfy the dispersion relation in the form derived above). 
Clearly, instability ($\omega^2<0$) is possible only if $\rho_1\Omega_1^2>\rho_2\Omega_2^2$.
This criterion is the discontinuous analogue of the condition $(\rho_0\Omega)'<\rho_0'g/\vvarpi$ derived in Section~\ref{secappinc}. It carries the signature of decreasing $\Omega^2$, as in MRI, but also of decreasing $\vvarpi^4\Omega^2$, as in CFI. In the discontinuous case these two contributions cannot be distinguished. In addition, when the densities differ, the criterion contains a buoyancy-like term: in the absence of gravity, where $g_{\rm eff}=-\Omega^2\vvarpi$, a positive density jump $\rho_1-\rho_2$ is destabilizing.  

To gain further physical insight, we examine several simple subcases in the following.

For the hydrodynamic case 
and same densities equation~(\ref{eqdispcentrrelapprox}) gives $ \sqrt{1-\dfrac{ 4\Omega_1^2 }{\omega^2} } - \sqrt{1-\dfrac{4\Omega_2^2 }{\omega^2} } =
\dfrac{4}{|k|\rin}$.
The requirement the left-hand side be positive means $\dfrac{ \Omega_2^2 }{\omega^2} >\dfrac{ \Omega_1^2 }{\omega^2} $
showing the influence of the angular momentum contrast in the centrifugal instability, since the sign of $\omega^2$ is the same with the sign of $\Omega_2^2-\Omega_1^2$. 
\\
For the hydrodynamic case with equal angular velocities, the approximate dispersion relation reduces to $\omega^2\sqrt{1-\dfrac{ 4\Omega^2 }{\omega^2} }=\dfrac{\rho_2-\rho_1}{\rho_2+\rho_1} \Omega^2\rin |k|$, showing explicitly the influence of the density contrast. In this limit the perturbation behaves as a Rayleigh-Taylor mode, with the role of gravity replaced by the centrifugal acceleration $\Omega^2\vvarpi\hat\vvarpi$.
\\
The stabilizing effect of the magnetic field (through its tension) becomes evident in the simplified case of equal densities and equal magnetic fields. In that case we obtain $\sqrt{1-\dfrac{ 4\omega^2\Omega_1^2 }{(k^2v_{\rm A}^2-\omega^2)^2} }-\sqrt{1-\dfrac{ 4\omega^2\Omega_2^2 }{(k^2v_{\rm A}^2-\omega^2)^2} }=\dfrac{4\omega^2}{(\omega^2-k^2v_{\rm A}^2)|k|\rin }$. 
Instability ($\omega^2<0$) is possible only when $\Omega_1^2>\Omega_2^2$.
After squaring and rearranging the above equation, one obtains a quadratic equation in $\omega^2$. 
Solving this quadratic and selecting the correct root (the one that satisfies the original dispersion relation) we obtain  
\begin{eqnarray}  
	\omega^2= 
	\dfrac{ -\dfrac{\Omega_1^2-\Omega_2^2 }{2 }k^2\rin^2}{
		\dfrac{2/{\cal A}+k^2\rin^2/M_{\rm A}^2}{1-k^2\rin^2/M_{\rm A}^4}
		+\sqrt{\left(\dfrac{2/{\cal A}+k^2\rin^2/M_{\rm A}^2}{1-k^2\rin^2/M_{\rm A}^4}\right)^2+\dfrac{k^2\rin^2-4}{1-k^2\rin^2/M_{\rm A}^4}}
	}
	\,, \label{eqdisrelsamefluids} 
	\\ 
	{\cal A}=\dfrac{\Omega_1^2-\Omega_2^2}{\Omega_1^2+\Omega_2^2}\,, \qquad 
	M_{\rm A}^2=\dfrac{\rho_1\Omega_1^2\rin^2-\rho_2\Omega_2^2\rin^2}{B_1^2+B_2^2}
	\,.
\end{eqnarray}
The above expression always gives real $\omega^2$: it is negative for $|k|\rin<M_{\rm A}^2$, but positive for $|k|\rin>M_{\rm A}^2$. 
Thus the magnetic field not only reduces the growth rate compared to the hydrodynamic case, but for sufficiently small wavelengths its tension becomes strong enough to completely stabilize the perturbation. The unstable regime corresponds to $|k|<\rho \dfrac{\Omega_1^2-\Omega_2^2}{2B^2}\rin$.

It is also instructive to derive an approximate dispersion relation by assuming that the perturbation is local, i.e. in the limit of large $k$ and $\rin$. 
In each region the radial acceleration satisfies Newton's law,
$\ddot \xi -\Omega^2(\vvarpi+\xi)=-\dfrac{\Pi_0'+\delta\Pi'}{\rho} - k^2v_{\rm A}^2\xi$, where the last term represents the magnetic-tension contribution. 
We assume $\delta\Pi'=i\tilde\kappa\delta\Pi$ with radial wavenumber $\tilde\kappa=\mp i |k|$, in analogy with the planar Rayleigh-Taylor instability in the incompressible limit (where $k_x^2+k_y^2+k_z^2=0$). The upper/lower sign corresponds to the inner/outer part, giving exponential decay of the amplitude away from the interface.
A more accurate approximation would use the WKBJ form of equation~(\ref{odefory2}), which also yields radial wavenumber $\mp i |k|$ in the limit of slow rotation.
Likewise, applying WKBJ to equation~(\ref{odefory1}) gives radial wavenumber $\mp i \Lambda_{1,2}$ from equation~(\ref{eqdispcentrtwo}), and in the slow-rotation limit $\Lambda_{1,2}\approx |k|$.
However, the goal here is not to repeat the full derivation, but to isolate the essential physics and obtain a compact approximate relation. 
\\ Substituting into Newton's law gives, in both regimes, $-\omega^2\xi-\Omega^2\xi=\mp |k|\dfrac{\delta\Pi}{\rho} - k^2v_{\rm A}^2\xi$. From this we obtain $\delta \Pi$, and hence the Lagrangian perturbation of the total pressure at the displaced position of each fluid element, $\Delta \Pi =\delta \Pi+\Pi_0'\xi\approx \mp\dfrac{k^2v_{\rm A}^2-\omega^2}{|k|}\rho\xi+\rho\Omega^2\vvarpi\xi$.
At the tangential discontinuity, both $\xi$ and $\Delta\Pi$ must be continuous, giving
$-\dfrac{k^2v_{\rm A 1}^2-\omega^2}{|k|}\rho_1+\rho_1\Omega_1^2\rin=
\dfrac{k^2v_{\rm A 2}^2-\omega^2}{|k|}\rho_2+\rho_2\Omega_2^2\rin$.
This reproduces the approximate dispersion relation~(\ref{eqdispcentrrelapprox}), with the square-root factors replaced by unity. The resulting expression is 
\begin{equation}
\omega^2=-\dfrac{\rho_1\Omega_1^2\rin-\rho_2\Omega_2^2\rin}{\rho_1+\rho_2}|k|+\dfrac{B_1^2+B_2^2}{\rho_1+\rho_2}k^2
\,,
\label{eqdispcentrplanar}
\end{equation}
which is essentially identical to the planar Rayleigh-Taylor dispersion relation (see \S~4 of Ref.~\cite{2025Symm...17..150V}), with the gravitational force densities $\rho_{1,2}g$ replaced by the centrifugal force densities $\rho_{1,2}\Omega_{1,2}^2\rin$.
This approximate expression predicts that $\omega^2<0$ for 
$0<|k|<\dfrac{M_{\rm A}^2}{\rin}$, and that the maximum growth rate is $\dfrac{M_{\rm A}}{2}\sqrt{\dfrac{\rho_1\Omega_1^2-\rho_2\Omega_2^2}{\rho_1+\rho_2}}$ at $k=\dfrac{M_{\rm A}^2}{2\rin}$.

\subsection{Application of WKBJ}\label{wkbjinterface}
The simplest variant of the WKBJ approach is to treat the array of the system~(\ref{eqsystemxxxyyy}) as approximately constant. In that case the system admits exponential solutions with a constant ratio $\xxx/\yyy$.  
Equivalently, the principal equation has approximately constant solutions. By dropping the $Y'$ term, these solutions are the roots of 
$f_{12}Y^2-2f_{11}Y-f_{12}=0 $, i.e., the $ \dfrac{1}{Y}=\dfrac{f_{11}\mp \sqrt{f_{11}^2+f_{12}f_{21}}}{-f_{12}}
= \rho\Omega^2 \pm \dfrac{\rho (k^2v_{\rm A}^2-\omega^2)\Lambda }{k^2\vvarpi} $ with $\Lambda=\sqrt{k^2-\dfrac{4\omega^2\Omega^2 k^2}{(k^2 v_{\rm A}^2-\omega^2)^2}}$.
The two signs correspond to $\dfrac{\xxx'}{\xxx}=\dfrac{\yyy'}{\yyy}=\mp\Lambda$, as follows directly from system~(\ref{eqsystemxxxyyy}).
Applying this in the two regions, the upper sign must be used for $\vvarpi>\rin$ and the lower sign for $\vvarpi<\rin$, ensuring exponential decay away from the interface on both sides. 
The dispersion relation follows from continuity of $Y$, giving
$\rho_1\Omega_1^2 - \dfrac{\rho_1 (k^2v_{\rm A 1}^2-\omega^2)\Lambda_1 }{k^2\rin}  =\rho_2\Omega_2^2 + \dfrac{\rho_2 (k^2v_{\rm A 2}^2-\omega^2)\Lambda_2 }{k^2\rin} $. This is identical to equation~(\ref{eqdispcentrrelapprox}). As expected, the WKBJ approach reproduces the large $|k|\rin $ limit of the exact dispersion relation.

\begin{figure}
	\includegraphics[width=.47\textwidth]{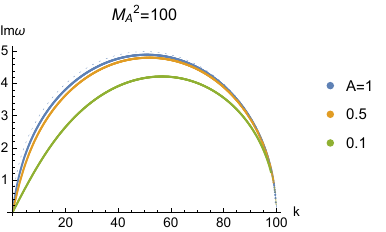} 
	\hfill
	\includegraphics[width=.47\textwidth]{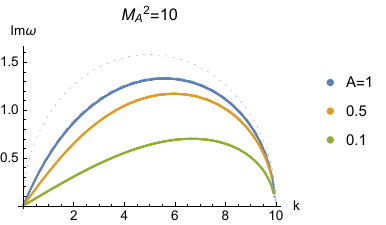} 
	
	\vspace{4mm}
	\includegraphics[width=.47\textwidth]{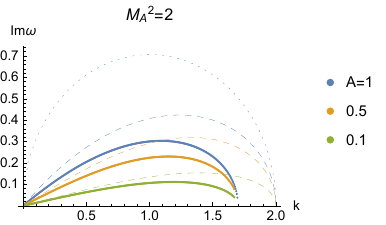}
	\hfill
	\includegraphics[width=.48\textwidth]{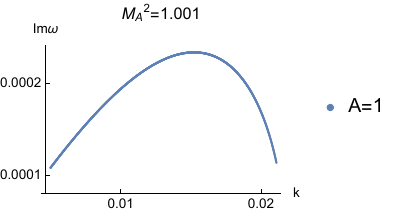}
	
	\vspace{3mm}
	\includegraphics[width=.33\textwidth]{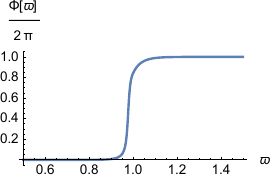}
	\includegraphics[width=.33\textwidth]{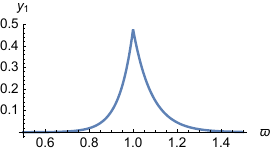}
	\includegraphics[width=.33\textwidth]{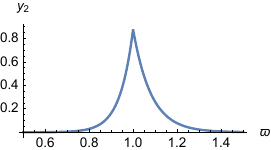}  
	\caption{Upper and middle rows: Growth rates of the instability for two incompressible fluids in contact, with equal density and equal magnetic field, rotating uniformly but with different angular velocities. Results are shown for several values of the angular velocity contrast measured through ${\cal A}=\dfrac{\Omega_1^2-\Omega_2^2}{\Omega_1^2+\Omega_2^2}$, and for several values of the magnetic field measured through $M_{\rm A}^2=\dfrac{\rho_1\Omega_1^2\rin^2-\rho_2\Omega_2^2\rin^2}{B_1^2+B_2^2}$. Solid curves correspond to the exact dispersion relation~(\ref{eqdispcentrrel}), dashed curves to the approximate expression~(\ref{eqdisrelsamefluids}), and dotted curves to the local planar-type approximation~(\ref{eqdispcentrplanar}). In all plots the wavenumber is given in units of $\dfrac{1}{\rin}$ and the growth rate in units of $\sqrt{\dfrac{\rho_1\Omega_1^2-\rho_2\Omega_2^2}{\rho_1+\rho_2}}$.
	Bottom row: Eigenfunctions for the case $M_A^2=100$, ${\cal A}=1$ and $k=10$.
	\label{figallMsq} }
\end{figure}

(We note that other variants of WKBJ exist, see Section 4 of Ref.~\cite{2023Univ....9..386V}.
The one used here is the ``zeroth-order WKBJ for the system''.) 

Summarizing, the exact dispersion relation for axisymmetric perturbations in the incompressible limit can be obtained either by numerical integration of the principal equation or analytically through equation~(\ref{eqdispcentrrel}). 
In the limit of large $k\rin $, the dispersion relation reduces to the approximate form~(\ref{eqdispcentrrelapprox}), which -- when the two sides contain identical fluids -- yields the explicit solution~(\ref{eqdisrelsamefluids}).
Furthermore, in the combined limit of large $k\rin $ and slow rotation, the local (planar-type) approximation leads to equation~(\ref{eqdispcentrplanar}).

Figure~\ref{figallMsq} shows results for the case of two fluids with the same density and the same magnetic field. This is purely a CFI associated with the decrease of $\Omega^2$ across the interface.
(Equivalently, one may call it MRI, since in this discontinuous configuration the criteria for CFI and MRI reduce to the same inequality $\Omega_1^2>\Omega_2^2$.) There is no buoyancy contribution because the density is uniform in the examined case.

\begin{figure}
	\includegraphics[width=.37\textwidth]{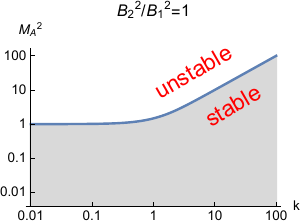} 
	\hfill
	\includegraphics[width=.37\textwidth]{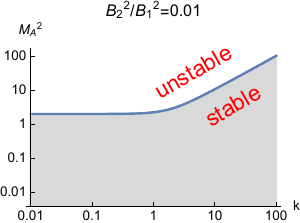} 	
	
	\vspace{5mm}
	\includegraphics[width=.37\textwidth]{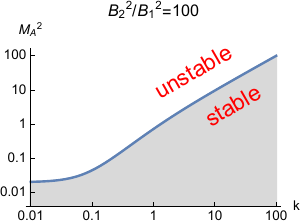}
	\hfill
	\includegraphics[width=.37\textwidth]{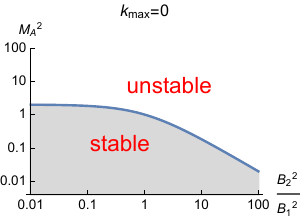} 
	\caption{Range of the CFI for two uniformly rotating incompressible fluids in contact. 
	The first three panels show marginal stability curves, i.e., $k_{\rm max}$ as a function of $M_{\rm A}^2=\dfrac{\rho_1\Omega_1^2\rin^2-\rho_2\Omega_2^2\rin^2}{B_1^2+B_2^2}$ for three values of $B_2^2/B_1^2$. 
	The bottom right panel indicates the parameter combinations for which the perturbation is steady at all wavelengths.
	\label{figallMsq2}
	}
\end{figure}

The exact region of instability -- i.e., the maximum wavenumber $k_{\rm max}$ below which $\omega^2<0$ -- is obtained by setting $\omega^2=0$ in equation~(\ref{eqdispcentrrel}). This yields 
\begin{eqnarray} 
M_{\rm A}^2 
= \dfrac{k_{\rm max}\rin}{2}\left[\dfrac{I_0(k_{\rm max}\rin)}{I_1(k_{\rm max}\rin)}+
\dfrac{K_0(k_{\rm max}\rin)}{ K_1(k_{\rm max}\rin)} \right] \,, 
\label{ma2kvarpi}
\end{eqnarray} 
a relation depending only on the parameter $M_{\rm A}^2$. The resulting marginal-stability curve is shown in the top-left panel of Figure~\ref{figallMsq2}. 
For relatively weak magnetic fields corresponding to large $M_{\rm A}^2$ and large $k_{\rm max}\rin$, the solution agrees with the estimation $k_{\rm max}\rin=M_{\rm A}^2$ discussed earlier. 
For strong magnetic fields this approximation breaks down because the assumption $k_{\rm max}\rin\gg 1$ no longer holds. The exact result is that $k_{\rm max}\rin$ approaches zero as $M_{\rm A}\to 1$, implying that for $M_{\rm A}< 1$ the perturbation is stable at all wavelengths.\endnote{
Equation~(\ref{ma2kvarpi}), evaluated in the limit where the argument of the modified Bessel functions $k_{\rm max}\rin$ is large and $I_0\approx I_1$, $K_0\approx K_1$, yields the expected relation $M_{\rm A}^2\approx k_{\rm max}\rin$.
In the opposite limit of small $k_{\rm max}\rin$, using the small-argument expansions $\dfrac{I_0(k_{\rm max}\rin)}{I_1(k_{\rm max}\rin)}\approx \dfrac{2}{k_{\rm max}\rin}$,
$ \dfrac{K_0(k_{\rm max}\rin)}{ K_1(k_{\rm max}\rin)}\to 0$, we obtain $M_{\rm A}^2\to 1$. 
} 

In the more general cases where the densities and magnetic fields differ between the two regions, the marginal stability curve also depends on the magnetic field contrast.
The maximum wavenumber is given by 
\begin{eqnarray}  
	\dfrac{\rho_1\Omega_1^2\rin^2-\rho_2\Omega_2^2\rin^2}{\rho_1v_{\rm A 1}^2+\rho_2v_{\rm A 2}^2}=
	\dfrac{B_1^2k_{\rm max}\rin}{B_1^2+B_2^2} \dfrac{I_0(k_{\rm max}\rin)}{I_1(k_{\rm max}\rin)} +\dfrac{B_2^2k_{\rm max}\rin}{B_1^2+B_2^2}\dfrac{K_0(k_{\rm max}\rin)}{ K_1(k_{\rm max}\rin)} 
	\,, \label{eqmarginaldifffluids} 
\end{eqnarray} 
which depends on two parameters: $M_{\rm A}^2$, and $B_2^2/B_1^2$. These dependencies are illustrated in the first three panels of Figure~\ref{figallMsq2}.
At large $k$ the result agrees with the approximate dispersion relation~(\ref{eqdispcentrplanar}), giving  
$k_{\rm max}\rin=M_{\rm A}^2$. At small $k$, however, the exact solution shows that $k_{\rm max}\rin$ vanishes at a finite value of $M_{\rm A}^2$, as displayed in the bottom-right panel of Figure~\ref{figallMsq2}.

\section{Compressibility Effects}\label{seccompres}

\subsection{Approximate Dispersion Relation}\label{secwkbjwallcompr}

The WKBJ approximation gives a compact way to understand how rotational instabilities are modified when the sound velocity is finite (and therefore $S$ is also finite). We focus on axisymmetric disturbances ($m=0$) in rotating flows where the unperturbed magnetic field is purely axial. 

The local wavenumber in the radial direction can be approximated by $\kappa_1$ from equation~(\ref{simplekappa1}), which in the present case becomes 
$ \kappa_1^2= \ktilde^2+\dfrac{\ktilde^2 (\vvarpi^4\Omega^2)'}{\vvarpi^3(k^2v_{\rm A}^2-\omega^2)}-\dfrac{4\ktilde^2 k^2v_{\rm A}^2\Omega^2 }{(k^2v_{\rm A}^2-\omega^2)^2}-\dfrac{\ktilde^2g_{\rm eff}\rho_0'}{\rho_0(k^2v_{\rm A}^2-\omega^2)} -\dfrac{k^2g_{\rm eff}^2}{S/\rho_0^2} -\left(\dfrac{f_{12}'}{2f_{12}}\right)^2 +\dfrac{f_{12}'}{ f_{12}}\dfrac{\omega^2 g_{\rm eff}}{S/\rho_0^2}+\left(\dfrac{f_{12}'}{2f_{12}}-\dfrac{\omega^2 g_{\rm eff}}{S/\rho_0^2}\right)'$.
Following the Boussinesq approximation and assuming $|k|\vvarpi \gg 1$, the last three terms become negligible.
Substituting $S/\rho_0^2=(c_s^2+v_{\rm A}^2)\omega^2-k^2v_{\rm A}^2c_s^2 $ and approximating $\tilde\kappa^2=\dfrac{\omega^4}{S/\rho_0^2}-k^2\approx -k^2$, which remains valid for large $|k|$ even when compressibility is included, provided that $|\omega^2|/k^2\ll c_s^2+v_{\rm A}^2$,  a condition expected to hold for the instability branches and verifiable a posteriori, the approximate dispersion relation becomes
\begin{adjustwidth}{-\extralength}{0cm}\begin{eqnarray} 	
	\dfrac{k_\vvarpi^2+k^2}{k^2}=
	-\dfrac{(\vvarpi^4\Omega^2)'}{\vvarpi^3(k^2v_{\rm A}^2-\omega^2)}
	+\dfrac{4k^2v_{\rm A}^2\Omega^2 }{(k^2v_{\rm A}^2-\omega^2)^2}
	+\dfrac{g_{\rm eff}\rho_0'}{\rho_0(k^2v_{\rm A}^2-\omega^2)} 
	+\dfrac{g_{\rm eff}^2}{k^2v_{\rm A}^2c_s^2-(c_s^2+v_{\rm A}^2)\omega^2}  
	\,. \label{eqmricompr}
\end{eqnarray}\end{adjustwidth}
This is a cubic with respect to $\omega^2$, but can be further simplified in various limits.

In the hydrodynamic case it gives 
$\omega^2=
\left[\dfrac{( \vvarpi^4\Omega^2)'}{ \vvarpi^3 } 
-\dfrac{g_{\rm eff}\rho_0'}{ \rho_0}
-\dfrac{g_{\rm eff}^2}{ c_s^2}\right]\dfrac{k^2}{k_\vvarpi^2+k^2}$, and we see that compressibility destabilizes.
Inside the bracket we recognize the square of the epicyclic frequency and the square of the effective buoyancy (or Brunt-V\"ais\"al\"a) frequency in the compressible limit $N_{\rm eff}^2=-\dfrac{g_{\rm eff}\rho_0'}{ \rho_0}
-\dfrac{g_{\rm eff}^2}{ c_s^2}$, with the last term corresponding to the variation of density of a fluid parcel as it evolves adiabatically in comparison with the variation of the density of its surroundings. 
So the result is a mixture of CFI and buoyancy instability in the compressible limit.  
Note that, using the equilibrium condition in the hydrodynamic case $P_0'=-\rho_0g_{\rm eff}$ we can write 
$N_{\rm eff}^2=-\dfrac{P_0'}{\rho_0}\left(\ln \dfrac{P_0^{1/\Gamma}}{\rho_0}\right)'
$, and thus buoyancy destabilizes when $\rho_0/P_0^{1/\Gamma}$ decreases in the direction of the effective gravity.
\\
For zero rotation we recover the known result $\omega^2=\dfrac{k^2N^2}{k_\vvarpi^2+k^2}$ with $N^2=-\dfrac{g\rho_0'}{ \rho_0}-\dfrac{g^2}{ c_s^2}=g \left(\ln \dfrac{P_0^{1/\Gamma}}{\rho_0}\right)'$, and the Schwarzschild instability criterion for convection $ \left(\ln \dfrac{P_0^{1/\Gamma}}{\rho_0}\right)'<0$.
\\
With rotation but without gravity (when the effective gravity is only the centrifugal $g_{\rm eff}=-\Omega^2\vvarpi$) we get 
$\omega^2=
\left[\dfrac{( \rho_0\vvarpi^4\Omega^2)'}{ \rho_0\vvarpi^3 } 
-\dfrac{\Omega^4\vvarpi^2}{ c_s^2}\right]\dfrac{k^2}{k_\vvarpi^2+k^2}$, a relation that gives the instability condition $\dfrac{d\ln(\rho_0\vvarpi^4\Omega^2)}{d\ln\vvarpi}<\dfrac{\Omega^2\vvarpi^2}{c_s^2}$ found in Ref.~\cite{2018MNRAS.475L.125G} using a heuristic analysis.
\\ In the more general hydrodynamic case with gravity included the instability condition becomes 
$N_{\rm eff}^2+(\vvarpi^4\Omega^2)'/\vvarpi^3<0$, or, 
$\dfrac{(\rho_0\vvarpi^4\Omega^2)'}{\rho_0\vvarpi^3 } 
-\dfrac{g\rho_0'}{ \rho_0}
<\dfrac{(g-\Omega^2\vvarpi)^2}{ c_s^2}$. Notably, it does not directly involve the Mach number $\Omega\vvarpi/c_s$, as in the case of zero gravity, but only the sum of the squares of the epicyclic frequency and of the effective buoyancy frequency.
The criterion can also be written as $\dfrac{(\vvarpi^4\Omega^2)'}{\vvarpi^3}-\dfrac{P_0'}{\rho_0}\left(\ln \dfrac{P_0^{1/\Gamma}}{\rho_0}\right)'<0$, generalizing the Schwarzschild instability criterion for convection in the presence of rotation.

With magnetic field included we can find a simple approximate dispersion relation in the limit of relatively weak magnetic field for which $k^2v_{\rm A}^2c_s^2-(c_s^2+v_{\rm A}^2)\omega^2\approx c_s^2(k^2v_{\rm A}^2-\omega^2)$. In that limit we get 
$\dfrac{k_\vvarpi^2+k^2}{k^2}(\omega^2-k^2v_{\rm A}^2)^2
-\left[\dfrac{(\vvarpi^4\Omega^2)'}{\vvarpi^3 }+N_{\rm eff}^2\right](\omega^2-k^2v_{\rm A}^2)
-4\Omega^2k^2v_{\rm A}^2 =0 $,
where 
\begin{eqnarray}
N_{\rm eff}^2=-\dfrac{g_{\rm eff}\rho_0'}{ \rho_0}
-\dfrac{g_{\rm eff}^2}{ c_s^2}
\,. \end{eqnarray}
Comparing with equation~(\ref{eqmri}) we see that only the full expression of the buoyancy frequency has replaced the one in the incompressible limit. 
Only the smallest root of the above quadratic can be negative, i.e., the
\begin{eqnarray}
\omega^2=k^2v_{\rm A}^2+\dfrac{N_{\rm eff}^2+(\vvarpi^4\Omega^2)'/\vvarpi^3}{2(1+k_\vvarpi^2/k^2)}
-\sqrt{\left[\dfrac{N_{\rm eff}^2+(\vvarpi^4\Omega^2)'/\vvarpi^3}{2(1+k_\vvarpi^2/k^2)}\right]^2+\dfrac{4\Omega^2 k^2v_{\rm A}^2}{1+k_\vvarpi^2/k^2}} \,.
\label{eqmricompr1}
\end{eqnarray}  
It is straightforward to show that this is indeed negative for wavenumbers 
$ k_\vvarpi^2+k^2<\dfrac{-N_{\rm eff}^2-\vvarpi(\Omega^2)'}{v_{\rm A}^2}$,
provided that the necessary condition for instability 
$N_{\rm eff}^2+\vvarpi(\Omega^2)'<0$, or, 
$\dfrac{\vvarpi(\rho_0\Omega^2)'}{\rho_0} 
-\dfrac{g\rho_0'}{ \rho_0}
<\dfrac{(g-\Omega^2\vvarpi)^2}{ c_s^2}$, holds. 

The fact that $N_{\rm eff}^2=-\dfrac{g_{\rm eff}\rho_0'}{ \rho_0}
-\dfrac{g_{\rm eff}^2}{ c_s^2}$ is smaller compared to its incompressible value $-\dfrac{g_{\rm eff}\rho_0'}{ \rho_0}$ and that $\omega^2$ is an increasing function of $N_{\rm eff}^2$ means that compressibility 
(i.e., to have finite $c_s$ in comparison with $c_s\to\infty$, while keeping everything else the same)
always destabilizes.

The resulting dispersion relation is a mixture of MRI corresponding to the differential rotation, and magnetic buoyancy (or Parker) instability \cite{1966ApJ...145..811P} corresponding to $N_{\rm eff}^2$, which indirectly depends on the magnetic field through the equilibrium condition $\Pi_0'=-\rho_0g_{\rm eff} $. 
The twin role of the magnetic field is evident in the instability condition as well as in the dispersion relation, since it appears both through $v_{\rm A}$, but also indirectly inside $N_{\rm eff}^2$.
The appearance of $v_{\rm A}$ is connected to the stabilizing nature of the magnetic tension and is responsible for the fact that unstable modes exist up to some maximum wavenumber. 
To see the effect of the magnetic field through the effective buoyancy frequency we need to think if and how it affects the density stratification by solving the equilibrium condition.  
For example in the subcase of a polytropic relation in the unperturbed state characterized by a constant $\Pi_0/\rho_0^{\Gamma_{\rm pol}}$ and constant index $\Gamma_{\rm pol}$ (as is usually assumed in studies of the Parker instability) the equilibrium condition $\Pi_0'=-\rho_0g_{\rm eff}$ gives $ \rho_0'=-\dfrac{\rho_0^2 g_{\rm eff}}{\Gamma_{\rm pol}\Pi_0}$, 
and we get $N_{\rm eff}^2=\dfrac{g_{\rm eff}^2}{c_s^2}\left(\dfrac{\Gamma/\Gamma_{\rm pol}}{1+B_0^2/2P_0}-1\right)$, 
showing the destabilization effect of the magnetic field (increasing $B_0^2$ decreases $N_{\rm eff}^2$).
Essentially, a strong magnetic field in the case of constant $\Pi_0/\rho_0^{\Gamma_{\rm pol}}$ decreases the effect of the stable stratification that corresponds to the positive term 
$-\dfrac{g_{\rm eff}\rho_0'}{ \rho_0}$ in $N_{\rm eff}^2$. 

Needless to say, there are other interesting cases that warrant further exploration.
For example in the absence of rotation equation~(\ref{eqmricompr}) gives
$\dfrac{k_\vvarpi^2+k^2}{k^2}(c_s^2+v_{\rm A}^2)(\omega^2-k^2v_{\rm A}^2)^2 +\left[(k_\vvarpi^2+k^2)v_{\rm A}^4+\dfrac{g\rho_0'}{\rho_0}(c_s^2+v_{\rm A}^2)+g^2\right](\omega^2-k^2v_{\rm A}^2)+\dfrac{g\rho_0'}{\rho_0}k^2v_{\rm A}^4=0$,
corresponding to the pure magnetic buoyancy (or Parker) instability, 
for which we find that only the smaller root of the quadratic can become negative,
$\omega^2=\dfrac{k^2v_{\rm A}^2c_s^2}{c_s^2+v_{\rm A}^2}
+ \dfrac{(k_\vvarpi^2+k^2)v_{\rm A}^4-\dfrac{g\rho_0'}{\rho_0 }(c_s^2+v_{\rm A}^2)-g^2}{2(c_s^2+v_{\rm A}^2)(1+k_\vvarpi^2/k^2)} 
\linebreak
-\sqrt{\left[ 
	\dfrac{(k_\vvarpi^2+k^2)v_{\rm A}^4-\dfrac{g\rho_0'}{\rho_0}(c_s^2+v_{\rm A}^2)-g^2}{2(c_s^2+v_{\rm A}^2)(1+k_\vvarpi^2/k^2)}\right]^2
	+\dfrac{g^2k^4v_{\rm A}^4}{(c_s^2+v_{\rm A}^2)^2(k_\vvarpi^2+k^2)}}$, 
provided that the instability condition $c_s^2\left[(k_\vvarpi^2+k^2)v_{\rm A}^2-\dfrac{g\rho_0'}{\rho_0}\right]<g^2$ is satisfied.
This instability condition in the subcase of constant $\Pi_0/\rho_0^{\Gamma_{\rm pol}}$ (with constant $\Gamma_{\rm pol}$), when the equilibrium $\Pi_0'=-\rho_0g$ gives $ \rho_0'=-\dfrac{\rho_0^2 g}{\Gamma_{\rm pol}\Pi_0}$, becomes $\dfrac{(k_\vvarpi^2+k^2)v_{\rm A}^2c_s^2}{g^2}<1-\dfrac{\Gamma P_0}{\Gamma_{\rm pol}\Pi_0} $,
and requires $ \dfrac{\Gamma}{\Gamma_{\rm pol}}<1+\dfrac{B_0^2}{2P_0}$, i.e., for given positive $\Gamma_{\rm pol}$ it requires sufficiently strong magnetic field corresponding to sufficiently small plasma beta (or, alternatively, thermodynamics corresponding to a sufficiently small index $\Gamma $), a known characteristic of the Parker instability.
\\
The limit of negligible thermal pressure is also an interesting case to examine, similarly to Refs.~\cite{2000ApJ...540..372K,2016MNRAS.462.3031B}. 
\\
It is also possible to study the general properties of the cubic dispersion relation~(\ref{eqmricompr}) without any further approximation (to alleviate the assumption of relatively weak magnetic field that we consider above in the rotating cases, but also include finite thermal pressure). 
Even more generally, one could include azimuthal magnetic field as in Ref.~\cite{2005ApJ...628..879P}, and also consider nonaxisymmetric perturbations ($m\neq 0)$, as in Ref.~\cite{2002ApJ...569L.121K}. These generalizations are beyond the scope of the present paper.

\subsection{Approximate Dispersion Relation for Interface-Driven Instability}
Similarly to the incompressible case (Section~\ref{wkbjinterface}) we can use the same variant of WKBJ to find the dispersion relation of an interface-driven instability. 
The constant solutions of the principal equation are 
$ \dfrac{1}{Y}=\dfrac{f_{11}\mp \sqrt{-\kappa_0^2}}{-f_{12}}$
with $\kappa_0^2=-f_{11}^2-f_{12}f_{21}$ given by equation~(\ref{simplekappa0}),
and correspond to $\dfrac{\xxx'}{\xxx}=\dfrac{\yyy'}{\yyy}=\mp\sqrt{-\kappa_0^2}$.
Applying the above in the two sides of the interface, the upper sign should be used in the region $\vvarpi>\rin$ and the lower sign in the region $\vvarpi<\rin$, such that we get exponential decrease as we move away from the interface in both sides. 
The dispersion relation comes from the continuity of $Y$.

Using the expression of $\kappa_0^2$ for axisymmetric disturbances ($m=0$) of rotating flows in which the unperturbed magnetic field has only $\hat z$ component and the gravity is zero, i.e.,
\begin{eqnarray}  
\kappa_0^2=-\dfrac{(v_{\rm A}^2-\omega^2/k^2)(c_s^2-\omega^2/k^2)\Lambda^2- \Omega^4\vvarpi^2}{v_{\rm A}^2c_s^2-(c_s^2+v_{\rm A}^2)\omega^2/k^2 }
\,, \quad \Lambda=\sqrt{k^2-\dfrac{4\omega^2\Omega^2k^2}{(k^2v_{\rm A}^2-\omega^2)^2}} \,,
\end{eqnarray}
we get  
\begin{eqnarray}  
\dfrac{1}{Y}
=\dfrac{\rho_0\Omega^2c_s^2}{c_s^2-\omega^2/k^2}\pm\dfrac{\rho_0}{\vvarpi}\sqrt{\left[(v_{\rm A}^2-\omega^2/k^2)\Lambda^2-\dfrac{\Omega^4\vvarpi^2}{c_s^2-\omega^2/k^2}\right] \left[v_{\rm A}^2-\dfrac{c_s^2\omega^2/k^2}{c_s^2-\omega^2/k^2}\right]} 
\,.
\end{eqnarray} 
Thus the dispersion relation is 
\begin{eqnarray}
\left[	\dfrac{\rho_0\Omega^2\vvarpi c_s^2}{c_s^2-\omega^2/k^2}
	-\rho_0\sqrt{\left[(v_{\rm A}^2-\omega^2/k^2)\Lambda^2-\dfrac{\Omega^4\vvarpi^2}{c_s^2-\omega^2/k^2}\right] \left[v_{\rm A}^2-\dfrac{c_s^2\omega^2/k^2}{c_s^2-\omega^2/k^2}\right]}
	\right]_{\vvarpi=\rin^-}
=\nonumber \\
\left[ \dfrac{\rho_0\Omega^2\vvarpi c_s^2}{c_s^2-\omega^2/k^2}
	+\rho_0\sqrt{\left[(v_{\rm A}^2-\omega^2/k^2)\Lambda^2-\dfrac{\Omega^4\vvarpi^2}{c_s^2-\omega^2/k^2}\right] \left[v_{\rm A}^2-\dfrac{c_s^2\omega^2/k^2}{c_s^2-\omega^2/k^2}\right]}
	\right]_{\vvarpi=\rin^+}
\,. \label{dispwkbjinterface}
\end{eqnarray} 
In most cases of interest in the limit of large $|k|\rin$ we may approximate $c_s^2-\omega^2/k^2\approx c_s^2 $, and $v_{\rm A}^2c_s^2-(c_s^2+v_{\rm A}^2)\omega^2/k^2\approx c_s^2 (v_{\rm A}^2-\omega^2/k^2)$, and the expression simplifies to 
\begin{eqnarray}  
&	\left[\rho_0\Omega^2\vvarpi
	-\rho_0\dfrac{k^2v_{\rm A}^2-\omega^2}{|k| }\sqrt{1
-\dfrac{4\Omega^2\omega^2}{(k^2v_{\rm A}^2-\omega^2)^2}
-\dfrac{\Omega^4\vvarpi^2}{c_s^2(k^2v_{\rm A}^2-\omega^2)} } 
	\right]_{\vvarpi=\rin^-}
&	= \nonumber \\ 
&
	\left[\rho_0\Omega^2\vvarpi
	+\rho_0\dfrac{k^2v_{\rm A}^2-\omega^2}{|k| }\sqrt{1
-\dfrac{4\Omega^2\omega^2}{(k^2v_{\rm A}^2-\omega^2)^2}
-\dfrac{\Omega^4\vvarpi^2}{c_s^2(k^2v_{\rm A}^2-\omega^2)} } 
	\right]_{\vvarpi=\rin^+}
& . \label{dispwkbjinterface1}
\end{eqnarray} 
A necessary condition for instability is clearly $\left[\rho_0\Omega^2 \right]_{\vvarpi=\rin^-}>
\left[\rho_0\Omega^2 \right]_{\vvarpi=\rin^+}$. 

\subsection{Exact Results}\label{seccompre_exact}

To examine how compressibility affects the growth rate of the rotational instability, we consider axisymmetric perturbations of a rotating flow, assuming magnetic field $B_0(\vvarpi)\hat z$ and zero gravity.
For this case \( f_{11}=\dfrac{k^2c_s^2\Omega^2 \vvarpi}{\left(c_s^2+v_{\rm A}^2\right)\omega^2-k^2c_s^2v_{\rm A}^2} \),  
\(f_{12} 
= \dfrac{(\omega^2-k^2c_s^2)\vvarpi/\rho_0}{(c_s^2+v_{\rm A}^2)\omega^2-k^2c_s^2v_{\rm A}^2}
\),  
\( f_{21}=-\dfrac{\rho_0(\omega^2-k^2v_{\rm A}^2)}{\vvarpi}
+\dfrac{4\rho_0\omega^2 \Omega^2/\vvarpi}{\omega^2-k^2v_{\rm A}^2} 
+\dfrac{k^2 \rho_0(c_s^2+v_{\rm A}^2)\Omega^4 \vvarpi}{\left(c_s^2+v_{\rm A}^2\right)\omega^2-k^2c_s^2v_{\rm A}^2}  \),
and the principal equation is  
\begin{adjustwidth}{-\extralength}{0cm}\begin{eqnarray}  
		\dfrac{d }{d\vvarpi } \left(\dfrac{1}{Y}\right) =\dfrac{(\omega^2-k^2c_s^2)\vvarpi/\rho_0}{(c_s^2+v_{\rm A}^2)\omega^2-k^2c_s^2v_{\rm A}^2}\left(\dfrac{1}{Y}+\dfrac{\rho_0 k^2c_s^2\Omega^2}{ \omega^2-k^2c_s^2 }\right)^2
		+\dfrac{\rho_0(\omega^2-k^2v_{\rm A}^2)}{\vvarpi}
		-\dfrac{4 \rho_0\omega^2 \Omega^2/\vvarpi
		}{\omega^2-k^2v_{\rm A}^2 } 
		-\dfrac{\rho_0k^2\Omega^4\vvarpi
		}{\omega^2-k^2c_s^2}
		\,.
		\label{eqprincipalrotshell}
\end{eqnarray}\end{adjustwidth} 
Compared to the incompressible case there is an additional difficulty connected to the requirement the total pressure 
$\Pi_0=\dfrac{\rho_0c_s^2}{\Gamma}+\dfrac{\rho_0v_{\rm A}^2}{2}$
of the unperturbed state to satisfy the
equilibrium condition $\Pi_0'= \rho_0\Omega^2 \vvarpi$.
Still, it is possible in some cases to find analytical expressions for the solution of the principal equation and the dispersion relation under certain assumptions, see, e.g.
Appendix~\ref{appendixkummer} for the case of uniform angular velocity, and Appendix~\ref{appendixbesel} for the case of uniform rotational velocity (in both cases constant $c_s$ and $v_{\rm A}$ are assumed). 
The numerical integration is straightforward though, and covers all cases. 

Here we consider two cases, extending previous results by taking compressibility into account.
In the first, we assume a rotating shell between two solid cylindrical boundaries, $\vvarpi_1<\vvarpi<\vvarpi_2 $, in order to explore how compressibility modifies the results of Section~\ref{sectwocylindinc}. 
\\ In the second case, we assume a uniformly rotating cylindrical flow in $\vvarpi<\rin$ surrounded by a static homogeneous plasma in $\vvarpi>\rin$, generalizing the results of Section~\ref{secincomprtwoflows} (for a non-rotating outer part, ${\cal A}=1$). 
The static environment is characterized by its constant density $\rho_2$, sound velocity $c_{s2}$ and Alfv\'en velocity $v_{\rm A 2}$ corresponding to uniform magnetic field $v_{\rm A 2}\sqrt{\rho_2}\hat z$. These need only to satisfy the pressure balance at the interface in the unperturbed state, i.e., continuity of $\Pi_0=\dfrac{\rho_2c_{s2}^2}{\Gamma_2}+\dfrac{\rho_2v_{\rm A}^2}{2}$. Here we choose same density and magnetic field in the two regions, same polytropic index $\Gamma=5/3$, and continuous $c_s$ at the interface.   

In both cases we keep the unperturbed state the same as in the incompressible analogue, with the only difference being that the sound velocity (and the thermal pressure) is now finite, and has the proper spatial dependence required for the equilibrium $\Pi_0'=\rho_0\Omega^2\vvarpi$ with $\Pi_0=\dfrac{\rho_0c_s^2}{\Gamma}+\dfrac{\rho_0v_{\rm A}^2}{2}$ to be satisfied.
\\ In particular, for the rotating shell 
we assume uniform density $\rho_0$, uniform magnetic field $v_{\rm A}\sqrt{\rho_0}\hat z$, and power-law rotation $\Omega=\Omega_2\left(\dfrac{\vvarpi}{\vvarpi_2}\right)^b$, in which case integration of the equilibrium condition gives 
$c_s=\sqrt{c_{s1}^2+\dfrac{\Gamma\Omega_2^2}{\vvarpi_2^{2b}} \dfrac{\vvarpi^{2b+2}-\vvarpi_1^{2b+2}}{2b+2}}$, where $c_{s_1}$ is the sound velocity at the inner radius $\vvarpi_1=\vvarpi_2-\Delta\vvarpi$.
\\
Similarly, for the rotating cylinder  
we assume uniform density $\rho_0$, uniform magnetic field $v_{\rm A}\sqrt{\rho_0}\hat z$, and constant $\Omega $, when the integration of the equilibrium condition gives $c_s=\sqrt{c_{s1}^2+\dfrac{\Gamma\Omega^2\vvarpi^2}{2}}$, where $c_{s_1}$ is the sound velocity at the symmetry axis. 

The principal equation can be integrated using the Schwarzian approach, and the eigenvalues are obtained from the requirement that $\dfrac{\Phi|_2-\Phi|_1}{2\pi}$
be an integer (with the subscripts 1 and 2 refer to the extreme values of $\vvarpi$ in the entire region occupied by plasma; $\vvarpi_1$ may be finite or zero, and $\vvarpi_2$ may be finite or infinity).

\begin{figure}
	\includegraphics[width=.45\textwidth]{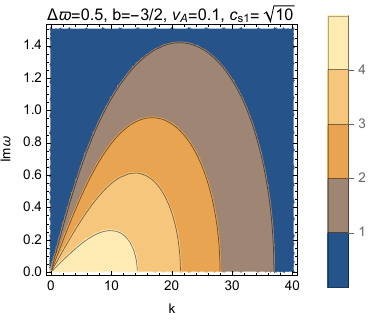}
	\hfill
	\includegraphics[width=.435\textwidth]{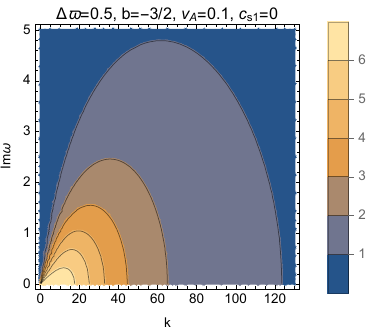} 
	\\
	\includegraphics[width=.45\textwidth]{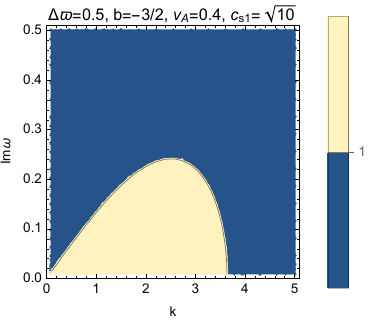}
	\hfill
	\includegraphics[width=.45\textwidth]{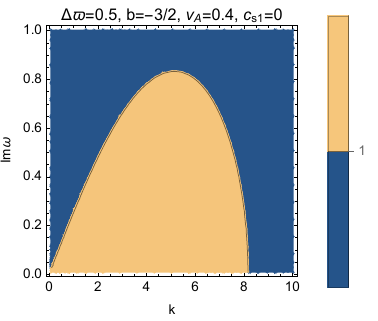}
	\caption{Effect of compressibility in the instability in a rotating shell with solid boundaries.
		As in Figure~\ref{figva01b1.5r05} we normalize all quantities using $\vvarpi_2$ as the unit length, and the angular velocity $\Omega|_{\vvarpi_2}$ as the unit frequency.
		\label{figva01cs1sq0b1.5r05}
	}
\end{figure}
\begin{figure}
	\includegraphics[width=.4\textwidth]{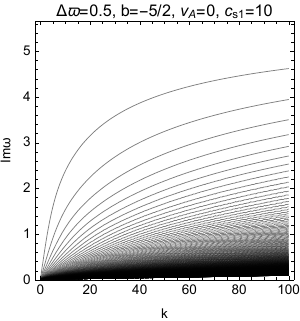}
	\hfill
	\includegraphics[width=.4\textwidth]{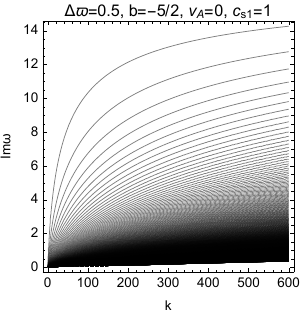}  
	\caption{Effect of compressibility on the CFI in a wall-bounded rotating hydrodynamic flow.
		\label{fighydrocomr}
	}
\end{figure}

Figure~\ref{figva01cs1sq0b1.5r05} shows how the growth rate changes with decreasing $c_s$ in the case of a wall-bounded rotational instability of a flow between two cylindrical walls.
In the top left panel the value of $c_{s1}$ is large enough and the growth rate is practically that of the incompressible limit 
(indeed, the top-left panel of Figure~\ref{figva01cs1sq0b1.5r05} is essentially identical to the top-left panel of Figure~\ref{figva01b1.5r05}).
The top-right panel of Figure~\ref{figva01cs1sq0b1.5r05} corresponds to the opposite extreme $c_{s1}=0$. 
\\ 
As expected from the WKBJ analysis of Section~\ref{secwkbjwallcompr}, compressibility is destabilizing, since it reduces the sum $N_{\rm eff}^2+(\vvarpi^4\Omega^2)'/\vvarpi^3$ appearing in the approximate dispersion relation~(\ref{eqmricompr1}) by a factor $\dfrac{g_{\rm eff}^2}{c_s^2}$, 
corresponding to a relative difference of the order $\dfrac{g_{\rm eff}^2/c_s^2}{\Omega^2}=\dfrac{\vvarpi^2\Omega^2}{c_s^2}$. 
Although the unperturbed density is uniform, buoyancy effects are present once compressibility is included, and they modify the dispersion relation accordingly. 
\\
The bottom panels of Figure~\ref{figva01cs1sq0b1.5r05} show the stabilization effect of the magnetic field.
In Section~\ref{secwkbjwallcompr} we discussed the twin role of the magnetic field, which stabilizes through its tension, but may destabilize if it affects the density stratification (as a consequence of the equilibrium condition). In the present case, however, the destabilizing effect is absent because the unperturbed density was chosen to be a constant.

Figure~\ref{fighydrocomr} shows the effect of compressibility in a hydrodynamic case. The left panel is practically identical to the corresponding panel of Figure~\ref{fighydroincomr}.
For smaller values of $c_{s1}$, however, the growth rate increases significantly. As in the MRI, compressibility destabilizes the CFI, and according to the WKBJ analysis of Section~\ref{secwkbjwallcompr}, the relative increase of the growth rate is expected to be of order $\sqrt{\dfrac{N_{\rm eff}^2}{(\vvarpi^4\Omega^2)'/\vvarpi^3}}=\dfrac{\vvarpi\Omega}{c_s}$. 

\begin{figure}
	\includegraphics[width=.45\textwidth]{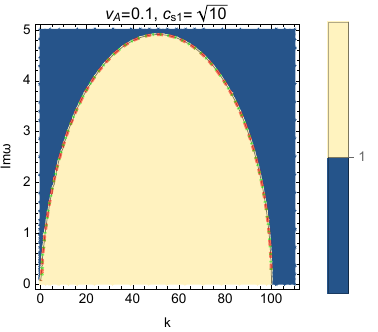}
	\hfill
	\includegraphics[width=.44\textwidth]{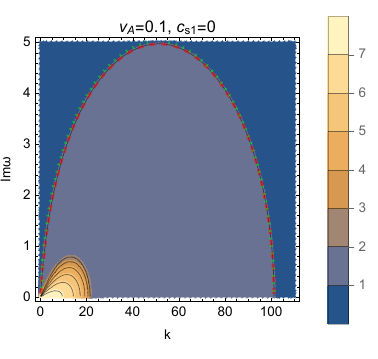}
	\\
	\includegraphics[width=.45\textwidth]{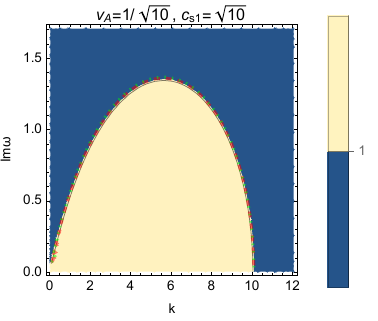}
	\hfill
	\includegraphics[width=.44\textwidth]{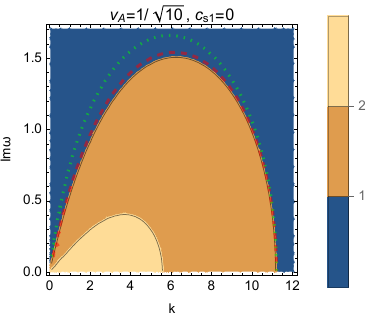}
	\\ \centerline{\includegraphics[width=.9\textwidth]{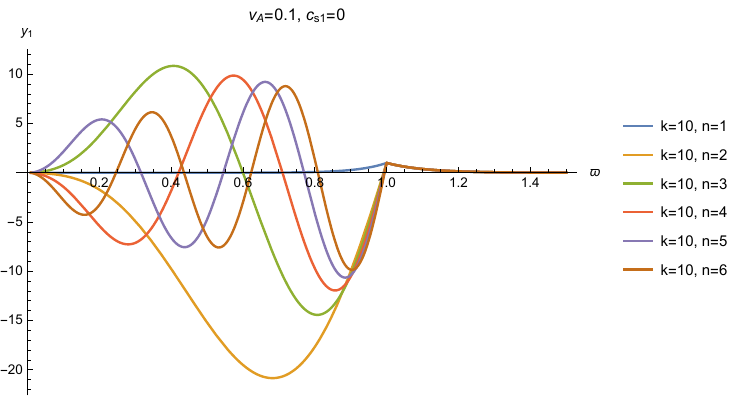}} 
	\caption{Effect of compressibility in the instability of a rotating cylinder surrounded by static plasma.
		The top row shows the dispersion relation for two values of the sound velocity on the axis, for Alfv\'en velocity $v_{\rm A}=0.1$ (corresponding to $M_{\rm A}^2=100$).
		The middle row shows the results for $v_{\rm A}=1/\sqrt{10}$ (corresponding to $M_{\rm A}^2=10$).
		The approximate dispersion relation~(\ref{dispwkbjinterface}) is also plotted with dashed red liens, and the approximate dispersion relation~(\ref{dispwkbjinterface1}) is plotted with dotted green liens. 
		Keeping the same normalization with Figure~\ref{figallMsq}, we use $\rin$ as the unit length, and $\sqrt{\dfrac{\Omega^2}{2}}$ as the unit frequency. 
		The bottom panel shows the eigenfunctions for the seven modes of the $c_{s1}=0$, $v_{\rm A}=0.1$ case for $k=10$ (normalized to their values at $\vvarpi=1$).
		In the region $\vvarpi>1$ all modes behave as $\sim e^{-k(\vvarpi-1)}$. In the region $\vvarpi<1$ the most unstable mode behaves as $\sim e^{k(\vvarpi-1)}$, while the remaining modes are oscillatory with radial wavenumber $k_\vvarpi\sim (n-1)\pi$. 
		\label{figva01cs1sq0b0r1}
	}
\end{figure}
Figure~\ref{figva01cs1sq0b0r1} shows how the MRI growth rate changes with decreasing $c_s$ in the case of a uniformly rotating cylinder surrounded by static plasma with same density and magnetic field.
In the left panels the value of $c_{s1}$ is large enough that the growth rate is essentially that of the incompressible limit 
(indeed, the top panels of Figure~\ref{figallMsq} -- the curves for ${\cal A}=1$ -- are practically identical to the left panels of Figure~\ref{figva01cs1sq0b0r1}).
The right panels of Figure~\ref{figva01cs1sq0b0r1} correspond to the opposite extreme $c_{s1}=0$. 
Similarly to the rotating shell, we observe a higher growth rate and a larger range of unstable wavenumbers due to the influence of buoyancy, although in this case the effect is less pronounced.
\\
For small enough $c_s$, additional modes are present, with distinctively different eigenfunctions, as seen in the bottom panel of Figure~\ref{figva01cs1sq0b0r1}. The CFI corresponds to the first mode with the larger growth rate, whose eigenfunction is localized around the tangential discontinuity (and becomes increasingly localized as $c_s$ decreases).  
\\ 
As already discussed, compressibility is destabilizing, but in this case the relative decrease of the sum $N_{\rm eff}^2+(\vvarpi^4\Omega^2)'/\vvarpi^3$ is of order $\dfrac{g_{\rm eff}^2/c_s^2}{4\Omega^2}=\dfrac{\rin^2\Omega^2}{4(c_{s1}^2+\Gamma\Omega^2\rin^2/2)}< \dfrac{1}{2\Gamma}$.
\\ The newly appearing modes ($n\ge 2$) are body modes, attributable to the magnetic buoyancy/Parker-type instability that arises once compressibility is included, even though the unperturbed density is uniform. 
They can be qualitatively understood using the WKBJ results of Section~\ref{secwkbjwallcompr}, in particular the dispersion relation~(\ref{eqmricompr1}) which becomes 
$\omega^2= k^2v_{\rm A}^2+\Omega^2\dfrac{4-\Omega^2\vvarpi^2/c_s^2}{2(1+k_\vvarpi^2/k^2)}
-\sqrt{\left[\Omega^2\dfrac{4-\Omega^2\vvarpi^2/c_s^2}{2(1+k_\vvarpi^2/k^2)}\right]^2+\dfrac{4 k^2v_{\rm A}^2 \Omega^2}{1+k_\vvarpi^2/k^2}} $,
and the instability condition $ k_\vvarpi^2+k^2<\dfrac{\Omega^4\vvarpi^2}{v_{\rm A}^2c_s^2 }$.  
The latter explains why these modes are absent in the incompressible limit, while in the opposite extreme $c_{s1}=0$ they are present whenever $ (n-1)^2\pi^2+k^2\rin^2<\dfrac{2\rin^2\Omega^2}{\Gamma v_{\rm A}^2}$.

\section{Summary}\label{secsummary}

The main goal of this paper is to present the form of the principal equation and the associated boundary conditions for hydrodynamic and ideal magnetohydrodynamic flows in cylindrical geometry. These results, which enable the application of the minimalist approach to any cylindrical equilibrium -- with radially varying density and pressure, axial and azimuthal components of both the velocity and magnetic field, and a radially directed gravitational field -- are summarized in Table~\ref{tab:formulas}. Within this framework, integrating the principal equation alone is sufficient to obtain the dispersion relation of linear perturbations.

We applied this methodology to rotating magnetized flows, primarily for illustration rather than to exhaust all possible equilibrium configurations. The purpose of these examples is not to introduce new physical instabilities, which are already well established, but to demonstrate how the minimalist approach provides a unified and analytically transparent framework for treating global cylindrical stability problems. The exact dispersion relations of the selected examples clearly reveal the underlying physical mechanisms, while the WKBJ approximation offers valuable insight. In particular, we derived approximate dispersion relations for axisymmetric perturbations in flows with axial magnetic fields, finding excellent agreement with exact solutions. These relations delineate the regimes in which centrifugal (CFI), magnetorotational (MRI), and buoyancy instabilities act individually or in combination, showing how the principal equation, boundary condition treatment, and Schwarzian approach together yield a compact and coherent description of these modes. 

CFI is governed by the epicyclic frequency, whose square is $(\vvarpi^4\Omega^2)'/\vvarpi^3$, and is active whenever this quantity is negative. In the presence of magnetic fields, CFI transitions to MRI, which is governed by the differential rotation frequency, $\vvarpi(\Omega^2)'$, and likewise operates when this quantity is negative. 
Buoyancy instability is controlled by the square of the effective buoyancy (or Brunt-V\"ais\"al\"a) frequency $N_{\rm eff}^2=-\dfrac{g_{\rm eff}\rho_0'}{\rho_0}-\dfrac{g_{\rm eff}^2}{c_s^2}$, where rotation modifies gravity through $g_{\rm eff}=g-\Omega^2\vvarpi$. Instability occurs when $N_{\rm eff}^2<0$. In the incompressible limit this reduces to the familiar condition of unstable density stratification, while compressibility introduces an additional negative term associated with adiabatic density changes of fluid parcels, further enhancing buoyancy-driven instability.  

In general, the stability of the system is determined by the sum of the squares of these characteristic frequencies; instability arises when the sum becomes negative.

Magnetic fields play a central role in the stability of rotating flows.
First, flux freezing couples neighboring fluid parcels along field lines, forcing them to rotate together and effectively replacing the epicyclic frequency by the differential rotation frequency in the stability criteria. This transforms CFI into MRI and typically has a destabilizing effect -- for example, even a weak magnetic field destabilizes an otherwise stable hydrodynamic Keplerian disk.
Second, magnetic tension stabilizes sufficiently short-wavelength perturbations, implying that any instability operates only up to a finite wavenumber $k_{\rm max}$.
Third, magnetic pressure can modify the equilibrium density stratification, altering the first term in $N_{\rm eff}^2$. This mechanism underlies magnetic-buoyancy (Parker) instability, where magnetic pressure weakens stabilizing stratification and promotes buoyant rise of magnetized regions. 

We examined two broad classes of flows: wall-bounded configurations and flows containing tangential discontinuities. Both are generally susceptible to instabilities -- global modes in the former and localized modes in the latter -- provided the relevant criteria are satisfied. In the presence of discontinuities, essentially the same instability criteria remain valid when spatial derivatives are interpreted as sharp gradients (i.e. jumps) across the interface. For example, a discontinuity in rotational velocity is associated with CFI/MRI and behaves like a local Rayleigh-Taylor instability for axisymmetric perturbations when the wavenumber is normal to the velocity shear, but may exhibit Kelvin-Helmholtz character when $m\neq 0$. 
Current‑driven and ballooning instabilities may also arise when azimuthal magnetic fields and appropriate pressure profiles are included.

The minimalist approach provides a unified and efficient way to obtain dispersion relations in all these situations, just as demonstrated in the examples of this paper. Furthermore, the alternative forms of the equations presented here may facilitate closed-form dispersion relations in additional cases.

Needless to say, although we selected rotational flows as the first applications of the minimalist approach in cylindrical geometry -- focusing on axisymmetric modes and axial magnetic fields -- the formalism summarized in Table~\ref{tab:formulas} can be applied to far more general cylindrical equilibria of disks and jets. These include equilibria with axial and azimuthal magnetic fields, rotational and axial velocity, and studies of nonaxisymmetric modes as well. We plan to explore these extensions in future works.

\vspace{6pt} 




\funding{This work was supported in part by the European Research Council through
	the Synergy Grant No.810218 (``The Whole Sun'', ERC-2018-SyG).}

\dataavailability{This research is analytical; no new data were generated or analyzed. If needed, more details on the study and the numerical results will be shared on reasonable request to the author.} 

\acknowledgments{The author would like to thank the anonymous referee for the careful and thorough reading of the manuscript, which also led to the identification of a typo in Appendix E.}

\conflictsofinterest{The author declares no conflicts of interest.} 


%


\appendixtitles{yes} 
\appendixstart
\appendix


\section{Linearization}\label{appendixA}

The linearized equations (\ref{mass}--\ref{induction}) are (for $\omega_0\neq 0$)
\begin{eqnarray}
	\omega_0 \dfrac{\Delta\rho}{\rho_0} =-i\dfrac{(\vvarpi V_{1\vvarpi})'}{\vvarpi} +kV_{1z}+\dfrac{m}{\vvarpi}V_{1\phi} \,,
	\label{mass1}
	\\
	P_1
	+i\dfrac{V_{1\vvarpi}}{\omega_0} P_0' -c_s^2\Delta\rho 
	=0 \,,
	\label{energy1}
	\\ -i\omega_0\rho_0V_{1\vvarpi}-\rho_1\dfrac{V_{0\phi}^2}{\vvarpi}-2\rho_0\dfrac{V_{0\phi}V_{1\phi}}{\vvarpi}=
	- \Pi_1' -2\dfrac{B_{0\phi}B_{1\phi}}{\vvarpi}+iFB_{1\vvarpi}
	-\rho_1  g \,,
	\label{momentumvarpi1}
	\\
	-i\omega_0\rho_0V_{1\phi}+\rho_0\dfrac{V_{1\vvarpi}}{\vvarpi} (\vvarpi V_{0\phi})' =
	-i\dfrac{m}{\vvarpi}\Pi_1+iFB_{1\phi}+\dfrac{B_{1\vvarpi}}{\vvarpi} (\vvarpi B_{0\phi})' \,, 
	\label{momentumphi1}
	\\
	-i\omega_0\rho_0V_{1z}+\rho_0 V_{1\vvarpi} V_{0z}'=
	-ik\Pi_1+iFB_{1z}+B_{1\vvarpi} B_{0z}' \,,
	\label{momentumz1}
	\\ 
	\omega_0 B_{1\vvarpi}= -F V_{1\vvarpi}
	\,, \label{inductionvarpi1}
	\\
	\omega_0 B_{1\phi}=\omega_0B_{0\phi}\dfrac{\Delta\rho}{\rho_0}
	-FV_{1\phi}  
	-i\vvarpi V_{1\vvarpi} \left(\dfrac{ B_{0\phi}}{\vvarpi } \right)'
	+i\vvarpi B_{1\vvarpi} \left(\dfrac{V_{0\phi}}{\vvarpi}\right)'
	\,, \label{inductionphi1}
	\\ 
	\omega_0 B_{1z}=\omega_0B_{0z}\dfrac{\Delta\rho}{\rho_0}-FV_{1z} 
	-i V_{1\vvarpi} B_{0z}' +iB_{1\vvarpi} V_{0z}'
	\,, \label{inductionz1}
\end{eqnarray} 
with $F=\bm k_0\cdot \bm B_0$, $c_s$ the sound velocity (only its unperturbed value is required and is assumed to be a known function of $\vvarpi$),
$P_0=\Pi_0-\dfrac{B_0^2}{2}$,
and $\Delta\rho=\rho_1+\dfrac{iV_{1\vvarpi}}{\omega_0} \rho_0'$.
Equation~(\ref{divB}) yields $ \dfrac{(\vvarpi B_{1\vvarpi})'}{\vvarpi} +i\dfrac{m}{\vvarpi}B_{1\phi}+ikB_{1z}=0$, a condition that is automatically satisfied when the above relations are used. 

We use equations (\ref{momentumphi1})--(\ref{inductionz1}),
together with the definitions $\Delta\rho=\rho_1+\dfrac{iV_{1\vvarpi}}{\omega_0}\rho_0'$, and $\xxx=\dfrac{i\vvarpi V_{1\vvarpi}}{\omega_0}$, to express the perturbations of the density, velocity and magnetic field in terms of $\Delta\rho$, $\xxx$, and $\Pi_1$: 
\begin{eqnarray}
	\rho_1=-\rho_0'\dfrac{\xxx}{\vvarpi}+\Delta\rho
	\,, \\
	V_{1\vvarpi} =\dfrac{\omega_0}{i\vvarpi}\xxx  \,, \\
	AV_{1\phi}=-A\left(\dfrac{V_{0\phi}}{\vvarpi}\right)'\xxx 
	-\dfrac{2\omega_0T }{\vvarpi^2}\xxx+
	\dfrac{m\omega_0}{\vvarpi}\Pi_1-F\omega_0B_{0\phi}\dfrac{\Delta\rho}{\rho_0}
	\,, \label{momentumphi2}
	\\
	AV_{1z}=-AV_{0z}'\dfrac{\xxx}{\vvarpi}  +
	k\omega_0\Pi_1-F\omega_0B_{0z}\dfrac{\Delta\rho}{\rho_0}
	\,, \label{momentumz2} \\
	B_{1\vvarpi}= iF \dfrac{\xxx}{\vvarpi}
	\,, \label{inductionvarpi2}
	\\
	A B_{1\phi}=-A \left(\dfrac{B_{0\phi}}{\vvarpi}\right)'\xxx 
	+2FT \dfrac{\xxx}{\vvarpi^2}-F
	\dfrac{m}{\vvarpi}\Pi_1 +\omega_0^2B_{0\phi}\Delta\rho
	\,, \label{inductionphi3}
	\\ 
	AB_{1z}=-AB_{0z}'\dfrac{\xxx}{\vvarpi} 
	-F k \Pi_1 +\omega_0^2B_{0z}\Delta\rho
	\,, \label{inductionz3}
\end{eqnarray} 
where $A=\rho_0\omega_0^2-F^2$, and $T=FB_{0\phi}+\rho_0\omega_0V_{0\phi}$.

The remaining equations (\ref{mass1})--(\ref{momentumvarpi1}) become
\begin{eqnarray}
	\vvarpi A\xxx'
	=(k^2\vvarpi^2+m^2)\Pi_1
	-2\dfrac{mT}{\vvarpi} \xxx-\omega_0^2 \vvarpi^2\Delta\rho
	  \,, \label{mass4}
	\\
	\omega_0^2\Pi_1+\omega_0^2\Pi_0'\dfrac{\xxx}{\vvarpi} = (Ac_s^2+\omega_0^2B_0^2)\dfrac{\Delta\rho}{\rho_0} -\left(F^2\vvarpi g-\W^2
	\right)\dfrac{\xxx}{\vvarpi^2}
	\,, 
	\label{energy4}
	\\ A\Pi_1'+A\xxx \left(\dfrac{\Pi_0'}{\vvarpi}\right)' 
	-\dfrac{\omega_0^2\vvarpi\Pi_0'+F^2\vvarpi g-\W^2}{\vvarpi}\Delta\rho
	\nonumber \\
	= A^2\dfrac{\xxx}{\vvarpi}-\dfrac{4T^2}{\vvarpi^3}\xxx 
	-A \rho_0 \left( \dfrac{g}{\vvarpi} \right)'\xxx
	+\dfrac{2mT\Pi_1}{\vvarpi^2} \,, \label{momentumvarpi4}
\end{eqnarray} 
where $ \W=\omega_0B_{0\phi}+FV_{0\phi} $, and we used $P_1=\Pi_1-B_{0z}B_{1z}-B_{0\phi}B_{1\phi}$, $P_0=\Pi_0-\dfrac{B_0^2}{2}$,
and the equilibrium of the unperturbed state $ \Pi_0'=\dfrac{\rho_0 V_{0\phi}^2}{\vvarpi}-\dfrac{ B_{0\phi}^2}{\vvarpi}-\rho_0  g $.

From now on we replace $\Pi_1=\yyy-\dfrac{\xxx}{\vvarpi}\Pi_0'$.
Equation (\ref{energy4}) can be solved for $\Delta\rho$:
\begin{eqnarray}
	\dfrac{S}{\rho_0^2}\Delta\rho = \dfrac{F^2\vvarpi g-\W^2}{\vvarpi^2}\xxx+\omega_0^2\yyy
	\,,
\end{eqnarray} 
where $S=\rho_0(Ac_s^2+\omega_0^2B_0^2) $.

Substituting into equations~(\ref{mass4}) and (\ref{momentumvarpi4}), we obtain the system~(\ref{eqsystemxxxyyy}) for $\xxx$ and $\yyy$. 

Once $\xxx$ and $\yyy$ are known from the solution of the system, the perturbation of every physical quantity can be found using 
\begin{eqnarray} 
	\rho_1=-\dfrac{\xxx}{\vvarpi}\rho_0'+\Delta\rho \,, \quad \Delta\rho=\dfrac{\rho_0^2}{S}\left(\omega_0^2\yyy+\dfrac{F^2\vvarpi g-\W^2}{\vvarpi^2}\xxx\right)
	\,, \label{energy6}
	\\ 
	V_{1\vvarpi}=-i\omega_0\dfrac{\xxx}{\vvarpi} \,, \label{v1varpi6} 
	\\ 
	V_{1\phi}=-\xxx \left(\dfrac{V_{0\phi}}{\vvarpi}\right)'+
	\dfrac{\omega_0}{\vvarpi A}\left(m\yyy-\dfrac{m\Pi_0'+2T}{\vvarpi}\xxx\right)-\dfrac{F \omega_0B_{0\phi}}{A\rho_0} \Delta\rho
	\,, \label{momentumphi6}
	\\
	V_{1z}=-\dfrac{\xxx}{\vvarpi}V_{0z}'  +
	\dfrac{k\omega_0}{A}\left(\yyy-\dfrac{\Pi_0'}{\vvarpi}\xxx\right)-\dfrac{F\omega_0B_{0z}}{A\rho_0} \Delta\rho
	\,, \label{momentumz6}
	\\
	B_{1\vvarpi}= iF \dfrac{\xxx}{\vvarpi}
	\,, \label{inductionvarpi6}
	\\
	B_{1\phi}=-\xxx \left(\dfrac{B_{0\phi}}{\vvarpi}\right)'
	-\dfrac{F}{\vvarpi A}\left(m\yyy-\dfrac{m\Pi_0'+2T}{\vvarpi}\xxx\right)+\dfrac{\omega_0^2B_{0\phi}}{A}\Delta\rho
	\,, \label{inductionphi6}
	\\ 
	B_{1z}=-\dfrac{\xxx}{\vvarpi} B_{0z}'
	-\dfrac{F k}{A}\left(\yyy-\dfrac{\Pi_0'}{\vvarpi}\xxx\right) +\dfrac{\omega_0^2B_{0z}}{A}\Delta\rho
	\,,\label{inductionz6} 
	\\ 
	P_1=-\dfrac{\xxx}{\vvarpi}P_0'+c_s^2 \Delta\rho
	\,. \label{pressure6}
\end{eqnarray}

\section{The Frieman-Rotenberg Formulation}\label{appendix_frieman} 
The Lagrangian perturbation of the velocity is related to the Lagrangian displacement vector through $\Delta\bm V= \dfrac{\partial\bm\xi}{\partial t}+(\bm V_0\cdot\nabla)\bm\xi=-i\omega_0\bm\xi-\xi_\phi\dfrac{V_{0\phi}}{\vvarpi}\hat\vvarpi
+\xi_\vvarpi\dfrac{V_{0\phi}}{\vvarpi}\hat\phi$ and the corresponding Eulerian perturbation is $\delta\bm V=\Delta\bm V-(\bm\xi\cdot\nabla)\bm V_0=\Delta\bm V+\xi_\phi\dfrac{V_{0\phi}}{\vvarpi}\hat\vvarpi-\xi_\vvarpi V_{0\phi}'\hat\phi -\xi_\vvarpi V_{0z}'\hat z$.
The Frieman-Rotenberg formulation is based on partially integrating the MHD equations so that all perturbations are expressed in terms of the Lagrangian displacement vector.
The continuity equation gives the Lagrangian perturbation of the density $\Delta\rho\, e^{i(m\phi+kz-\omega t)}=-\rho_0\nabla\cdot\bm\xi$ and the corresponding Eulerian perturbation is $\delta\rho=\Delta\rho\, e^{i(m\phi+kz-\omega t)}-\bm\xi\cdot\nabla\rho_0$. 
The induction equation gives for the Lagrangian perturbation of the magnetic field $\Delta\bm B=(\bm B_0\cdot\nabla)\bm\xi-\bm B_0(\nabla\cdot\bm\xi)=iF\bm\xi-\xi_\phi\dfrac{B_{0\phi}}{\vvarpi}\hat\vvarpi
+\xi_\vvarpi\dfrac{B_{0\phi}}{\vvarpi}\hat\phi +\dfrac{\Delta\rho}{\rho_0}\bm B_0$ and the corresponding Eulerian perturbation is $\delta\bm B=\nabla\times(\bm\xi\times\bm B_0)=\Delta\bm B-(\bm\xi\cdot\nabla)\bm B_0=\Delta\bm B+\xi_\phi\dfrac{B_{0\phi}}{\vvarpi}\hat\vvarpi-\xi_\vvarpi B_{0\phi}'\hat\phi -\xi_\vvarpi B_{0z}'\hat z$.
The Lagrangian perturbation of the total pressure is $\Delta\Pi=\Delta P+\bm B_0\cdot \Delta\bm B$ with the thermal part $\Delta P=c_s^2\Delta\rho\, e^{i(m\phi+kz-\omega t)}$ obtained from integrating the energy equation. The corresponding Eulerian perturbation is $\delta\Pi=\Delta\Pi-\bm\xi\cdot\nabla\Pi_0$. 
Thus, all Eulerian perturbations can be written as functions of the displacement
\begin{eqnarray}
\bm V_1=-i\omega_0\bm\xi_1-\xi_{1\vvarpi}V_{0z}'\hat z-\xi_{1\vvarpi}\vvarpi \left(\dfrac{V_{0\phi}}{\vvarpi}\right)' \hat \phi  \,,  
\label{V1frieman}
\\ 
\rho_1 +\xi_{1\vvarpi}\rho_0'=\Delta\rho=-\rho_0\left[i\bm k_0\cdot\bm \xi_1+\dfrac{(\vvarpi \xi_{1\vvarpi})'}{\vvarpi}\right] \,, 
\label{rho1frieman}
\\ 
\bm B_1 =iF \bm \xi_1 
+\dfrac{\Delta\rho}{\rho_0}\bm B_0 -\xi_{1\vvarpi}B_{0z}'\hat z-\xi_{1\vvarpi}\vvarpi \left(\dfrac{B_{0\phi}}{\vvarpi}\right)' \hat \phi   
\,, 
\label{B1frieman}
\\
\Pi_1
= iF  \bm B_0\cdot \bm \xi_1
+\xi_{1\vvarpi} \dfrac{B_{0\phi}^2}{\vvarpi } - \xi_{1\vvarpi}\Pi_0' 
+(\rho_0c_s^2+B_0^2)\dfrac{\Delta\rho}{\rho_0} 
\label{Pi1frieman}
\,.
\end{eqnarray}  
The displacement itself is determined by the three components of the momentum equation
\begin{eqnarray}  
	-A\bm \xi_1 +i\dfrac{2T}{\vvarpi} \left(\xi_{1\phi}\hat\vvarpi-\xi_{1\vvarpi}\hat\phi \right) =
	iF\dfrac{\Delta\rho}{\rho_0}\bm B_0
	-i \Pi_1 \bm k_0  
	\nonumber \\
	-\left(\Pi_1+\xi_{1\vvarpi}\Pi_0'\right)' \hat\vvarpi  
	-i\bm k_0\cdot\bm \xi_1 \Pi_0' \hat\vvarpi  
	-\rho_0\left(\dfrac{ g}{\vvarpi} \right)'\vvarpi\xi_{1\vvarpi}\hat\vvarpi 
	-\dfrac{\Delta\rho}{\rho_0} \dfrac{B_{0\phi}^2}{\vvarpi} \hat\vvarpi \,.
	\label{momentumallf}
\end{eqnarray}  
Notably, the eigenvalue $\omega$ appears only inside $A$ and $T$ on the left-hand side of the above equation.
The components $\xi_{1\phi}$ and $\xi_{1\phi}$ enter without derivatives and can be obtained directly from the $\hat\phi$ and $\hat z$ components of equation~(\ref{momentumallf}). 
Substituting these expressions into the $\hat \vvarpi$ component of equation~(\ref{momentumallf}) yields a second-order differential equation for $\xi_{1\vvarpi}$.

Instead of $\xi_{1\phi}$ and $\xi_{1z}$ we may equivalently use $\yyy$ and $\Delta\rho$ as in the previous sections. In this formulation $\xxx$, $\yyy$ satisfy the system~(\ref{eqsystemxxxyyy}), while $\Delta\rho$ is given by equation~(\ref{energy6}).

The two formulations are connected through the vector expression 
\begin{eqnarray}  
\bm \xi_1 = \dfrac{\xxx}{\vvarpi} \hat\vvarpi 
-i\dfrac{ 2 T+m\Pi_0'}{A }\dfrac{\xxx}{\vvarpi^2}\hat\phi
- i \dfrac{k\Pi_0' }{A}\dfrac{\xxx}{\vvarpi}\hat z 
+ i\dfrac{\yyy}{A}\bm k_0
-i\dfrac{F}{A }\dfrac{\Delta\rho}{\rho_0}\bm B_0 \,,
\end{eqnarray}   
or equivalently by the relations
\begin{eqnarray}  
\xxx=\vvarpi\xi_{1\vvarpi} \,, \quad 
\yyy=\xi_{1\vvarpi}\Pi_0'+\dfrac{(iA\xi_{1\phi}-2\xi_{1\vvarpi}T/\vvarpi)B_{0z}-iA\xi_{1z}B_{0\phi}}{B_{0\phi}k-B_{0z}m/\vvarpi} \,, \\ 
\dfrac{\Delta\rho}{\rho_0}=\dfrac{(iA\xi_{1\phi}-2\xi_{1\vvarpi} T/\vvarpi)k-iA\xi_{1z}m/\vvarpi}{F(B_{0\phi}k-B_{0z}m/\vvarpi)}
\,.
\end{eqnarray}

\section{Alternative Forms of the Equations}\label{appendix-sl-oscil}

\subsection{Alternative Forms of the Principal Equation}
The following alternative forms of the principal equation may be useful, particularly when seeking analytical solutions\endnote{The differential equation
	$xy'= m^2-x^2-y^2 \Leftrightarrow 
	x\left(\dfrac{1}{y}\right)'=\dfrac{x^2 }{y^2}+1-\dfrac{ m^2}{y^2}$  
	(which is transformed to Bessel $x^2B''+xB'+(x^2-m^2)B=0$ with $y=x\dfrac{B'}{B}$) has the general solution $y=|m|-x\dfrac{J_{|m|+1}(x)+CY_{|m|+1}(x)}{J_{|m|}(x)+CY_{|m|}(x)}$.
	Similarly, the differential equation 
	$x\left(\dfrac{1}{y}\right)'=\dfrac{x^2}{y^2}+1-\dfrac{ m^2-1}{x^2}$  
	(which is transformed to Bessel $x^2B''+xB'+(x^2-m^2)B=0$ with $\dfrac{1}{y}=-\dfrac{B+xB'}{x^2 B}$) has the general solution $\dfrac{1}{y}=-\dfrac{|m|+1}{x^2}+\dfrac{1}{x}\dfrac{J_{|m|+1}(x)+CY_{|m|+1}(x)}{J_{|m|}(x)+CY_{|m|}(x)}$.
	Another interesting case is the differential equation 
	$x\left(\dfrac{1}{y}\right)'=\dfrac{x^2}{y^2}-\dfrac{1}{4}-\dfrac{\alpha-3/2}{x}$, which is transformed to the Kummer equation $xU''+(3-x)U'-\alpha U=0$ with the substitution $\dfrac{1}{y}=\dfrac{1}{2x}-\dfrac{2}{x^2}-\dfrac{U'}{xU}$.
} 
\begin{eqnarray} 
	\left(\dfrac{1}{Y}\right)' =f_{12} \left(\dfrac{1}{Y}\right)^2+2f_{11}\left(\dfrac{1}{Y}\right)-f_{21}
\Leftrightarrow \nonumber  \\
	\left(\dfrac{f_{12}}{Y}+f_{11}+\dfrac{f_{12}'}{2f_{12}}\right)' = \left(\dfrac{f_{12}}{Y}+f_{11}+\dfrac{f_{12}'}{2f_{12}}\right)^2 +\kappa_1^2 \,,
\end{eqnarray}	
\begin{eqnarray}
	\left(f_{21} Y-f_{11}+\dfrac{f_{21}'}{2f_{21}}\right)' = \left(f_{21}Y-f_{11}+\dfrac{f_{21}'}{2f_{21}}\right)^2+\kappa_2^2 \,,  
\end{eqnarray} with 
\begin{eqnarray} 
	\kappa_1^2
	=\kappa_0^2+f_{12}\left(\dfrac{f_{11}}{f_{12}}\right)'-\left(\dfrac{f_{12}'}{2f_{12}}\right)^2+\left(\dfrac{f_{12}'}{2f_{12}}\right)'
	\,,\label{kappa1oscillator}
\\
	\kappa_2^2
=\kappa_0^2-f_{21}\left(\dfrac{f_{11}}{f_{21}}\right)'-\left( \dfrac{f_{21}'}{2f_{21}}\right)^2+\left(\dfrac{f_{21}'}{2f_{21}}\right)'
	\,,\label{kappa2oscillator}
\\
\kappa_0^2=-f_{11}^2-f_{12}f_{21}
 \,. \label{kappa0oscillator}
\end{eqnarray} 
After substituting the expressions of $f_{ij}$ we can write $\kappa_0^2$ in the form 
\begin{eqnarray}  
\kappa_0^2=\ktilde^2-\dfrac{\ktilde^2\rho_0\vvarpi}{A}\left(\dfrac{ g }{\vvarpi}\right)'
+\dfrac{4k^2T^2}{\vvarpi^2A^2} 
\nonumber \\ 
-\dfrac{k^2 }{ S/\rho_0^2} \left(g_{\rm eff}+\dfrac{2\omega_0 B_{0\phi} W}{\vvarpi A }\right)^2 
-\dfrac{1}{S/\rho_0^2} \left[\dfrac{2T\omega_0^2}{\vvarpi A}-\dfrac{m}{\vvarpi}\left(g_{\rm eff}+\dfrac{2\omega_0B_{0\phi} W}{\vvarpi A }\right)\right]^2
\,, \label{simplekappa0}
\end{eqnarray}
and the expression of $\kappa_1^2$ as
\begin{eqnarray} 	
\kappa_1^2= \ktilde^2 
+\dfrac{\ktilde^2\vvarpi}{A}\left(\dfrac{B_{0\phi}^2}{\vvarpi^2}\right)'-\dfrac{4\ktilde^2 FB_{0\phi}T}{\vvarpi^2A^2}  
+\dfrac{\ktilde^2g_{\rm eff}\rho_0'}{ A} -\dfrac{k^2+m^2/\vvarpi^2}{S/\rho_0^2 } \left(g_{\rm eff}+\dfrac{2\omega_0 B_{0\phi} W}{\vvarpi A }\right)^2
\nonumber	\\
-\dfrac{\ktilde^2\rho_0 (\vvarpi^2V_{0\phi}^2)'}{A\vvarpi^3}-\dfrac{4\ktilde^2 \rho_0FV_{0\phi} W}{\vvarpi^2A^2}
+\dfrac{2\omega_0^2}{S/\rho_0^2}\left(g_{\rm eff}+\dfrac{2\omega_0B_{0\phi} W}{\vvarpi A }\right)
\left( \dfrac{ f_{12}'}{2f_{12}}+ \dfrac{2m T}{\vvarpi^2 A}\right) 
\nonumber	\\
-\left(\dfrac{f_{12}'}{2f_{12}}+\dfrac{2mT}{\vvarpi^2A}\right)^2
+\left[\dfrac{f_{12}'}{2f_{12}}+\dfrac{2mT}{\vvarpi^2A}-\dfrac{\omega_0^2}{S/\rho_0^2}\left(g_{\rm eff}+\dfrac{2\omega_0B_{0\phi} W}{\vvarpi A }\right)\right]'
\,. \label{simplekappa1}
\end{eqnarray}  

\subsection{Sturm-Liouville Formulation}
We can reformulate the system (\ref{eqsystemxxxyyy}) of the two first-order differential equations for $\xxx$ and $\yyy$, into a single second-order differential equation for either $\xxx$ or $\yyy$, and obtain $Y$ afterwards: 
\begin{eqnarray}
	\left(\dfrac{\xxx'}{f_{12}}\right)'
	+\left[-f_{11}^2-f_{12} f_{21} +f_{12}\left(\dfrac{f_{11}}{f_{12}}\right)'\right]\dfrac{\xxx}{f_{12}} =0
	\,, \quad 
	\dfrac{1}{Y}=\dfrac{-f_{11}-\xxx'/\xxx }{f_{12}}
	\,,
	\label{odefory1}
\end{eqnarray}
\begin{eqnarray}
	\left(\dfrac{\yyy'}{f_{21}}\right)' 
	+\left[-f_{11}^2-f_{12} f_{21}
	-f_{21}\left(\dfrac{f_{11}}{f_{21}}\right)'\right]\dfrac{\yyy}{f_{21}} =0
	\,, \quad 
	Y=\dfrac{ f_{11}-\yyy'/\yyy}{f_{21}}
	\,. \label{odefory2}\end{eqnarray}

\subsection{``Variable Frequency Oscillator'' Formulation}\label{appvfo}
As explained in Ref.~\cite{2024Symm...16..648V} we can also reformulate the above as “variable
frequency oscillator” equations 
\begin{eqnarray}
	\left(\dfrac{\xxx}{\sqrt{f_{12}}}\right)''+\kappa_1^2 \dfrac{\xxx}{\sqrt{f_{12}}} =0 \,, \quad 
	\dfrac{1}{Y}=\dfrac{-f_{11}-\xxx'/\xxx }{f_{12}}
	\,, \label{odefory1oscillator}
\end{eqnarray}
\begin{eqnarray}
	\left(\dfrac{\yyy}{\sqrt{f_{21}}}\right)''+\kappa_2^2 \dfrac{\yyy}{\sqrt{f_{21}}} =0 \,, \quad 
	Y=\dfrac{ f_{11}-\yyy'/\yyy}{f_{21}}
	\,.\label{odefory2oscillator}
\end{eqnarray}

The above expressions give directly the local radial wavenumber $k_\vvarpi$ in the WKBJ approximation.
The corresponding approximate dispersion relation can therefore be obtained either from equation~(\ref{kappa1oscillator}) by setting $\kappa_1=k_\vvarpi$, or from equation~(\ref{kappa2oscillator}) by setting $\kappa_2=k_\vvarpi$. Notably, these two choices lead to distinct WKBJ approximations of the dispersion relation; see the related discussion in \S 4 of Ref.~\cite{2023Univ....9..386V}. As explained there, a third option is to apply the WKBJ method to the system associated with $\kappa_0=k_\vvarpi$.
The appropriate choice among $\kappa_1$, $\kappa_2$, and $\kappa_0$ depends on the relative smallness of the quantities $\kappa_1'/\kappa_1^2$, $\kappa_2'/\kappa_2^2$, and $\kappa_0'/\kappa_0^2$, respectively.

\section{Analytical Solution for Rotating Magnetized Flow with Constant $\Omega$}\label{appendixkummer}

We can analytically solve the principal equation~(\ref{eqprincipalrotshell}) for axisymmetric disturbances when $\Omega =$ constant, assuming also that the sound and Alfv\'en velocities are constant.
The equilibrium condition of the unperturbed state $\Pi_0'= \rho_0\Omega^2 \vvarpi$,
writing $\Pi_0=\dfrac{\rho_0\rin^2\Omega^2}{a} $ using the definitions of the sound and Alfv\'en velocities
($c_s^2= {\Gamma P_0}/{\rho_0}$, $v_{\rm A}^2= {B_0^2}/{\rho_0}$).
and defining the dimensionless parameter $a=\dfrac{\rin^2\Omega^2}{c_s^2/\Gamma+v_{\rm A}^2/2}$ with some reference length $\rin$,
yields the radial dependence of the density $\rho_0= \rhoin\exp\left\{\dfrac{a}{2}\left[\left(\dfrac{\vvarpi}{\rin}\right)^{2}-1\right]\right\}$, and consequently of all the other equilibrium quantities 
$\Pi_0=\dfrac{\rho_0\rin^2\Omega^2}{a}$,
$P_0=\dfrac{\rho_0c_s^2}{\Gamma}$, 
$\bm B_0=v_{\rm A}\sqrt{\rho_0}\hat z$, $\bm V_0=\vvarpi\Omega\hat\phi$.  

One way to obtain the analytical solution is to use the corresponding Sturm-Liouville formulation~(\ref{odefory1})
\begin{eqnarray}
	\left(\dfrac{\xxx'}{\vvarpi} \right)'
	+\dfrac{a\xxx'}{\rin^2} 
	+\left(\dfrac{ \lambda^2\Lambda^2 }{ k^2 }  
- a' \dfrac{\vvarpi^2}{\rin^4}
	\right)  \dfrac{\xxx}{\vvarpi} =0
	\,,  
	\quad \dfrac{1 }{Y}=\dfrac{\rho_0 k^2c_s^2\Omega^2}{k^2c_s^2-\omega^2}+
	\dfrac{\rho_0(k^2v_{\rm A}^2-\omega^2)\xxx'}{\lambda^2\vvarpi\xxx}
	\\
a=\dfrac{\rin^2\Omega^2}{c_s^2/\Gamma+v_{\rm A}^2/2}
\,, \quad 
a'
= \dfrac{-\lambda^2 k^2 a^2
[2(1-1/\Gamma)c_s^2-v_{\rm A}^2 ](2c_s^2/\Gamma+v_{\rm A}^2 )
}{4(k^2v_{\rm A}^2-\omega^2)(k^2c_s^2-\omega^2)} 
\,, 
\nonumber	\\
\lambda^2=-\dfrac{(k^2v_{\rm A}^2-\omega^2)(k^2c_s^2-\omega^2)}{k^2c_s^2v_{\rm A}^2-(c_s^2+v_{\rm A}^2)\omega^2}\,, \quad 
\Lambda=\sqrt{ k^2 -\dfrac{4\omega^2 \Omega^2k^2  }{(k^2v_{\rm A}^2-\omega^2)^2} }
	\,. \nonumber
\end{eqnarray} 
The equation for $\xxx$ can be solved analytically with the substitutions 
\\
${\rm z}=\sqrt{\dfrac{a^2}{4}+a'}\, \dfrac{\vvarpi^2}{\rin^2}$ and $\xxx=\dfrac{{\rm z} \, U({\rm z})}{\sqrt{\rho_0 e^{\rm z}}}$, 
when we find that $U$ satisfies the Kummer equation 
$ {\rm z}\dfrac{d^2U}{d{\rm z}^2}+(2-{\rm z})\dfrac{dU}{d{\rm z}}-\upalpha U=0$ with 
$\upalpha=1-\dfrac{ \lambda^2 \Lambda^2\rin^2}{2k^2\sqrt{a^2+4a'}}
$.

The general solution is a linear combination of the confluent hypergeometric function of the first kind $M(\upalpha,2,{\rm z})$ and the Tricomi's function $U(\upalpha,2,{\rm z})$, thus $U=C_1M(\upalpha,2,{\rm z})+C_2U(\upalpha,2,{\rm z})$
and 
$\xxx={\rm z}\dfrac{C_1M(\upalpha,2,{\rm z})+C_2U(\upalpha,2,{\rm z})}{\sqrt{\rho_0 e^{\rm z}}}$.
If the physical domain of interest includes the symmetry axis, only the solution proportional to $M(\upalpha,2,{\rm z})$ should be retained, while if the domain extends to large radii, the Tricomi function  $U(\upalpha,2,{\rm z})$ is the appropriate choice
(although, strictly speaking, a flow with uniform angular velocity extending to infinity is unphysical, as it implies both diverging density and unbounded azimuthal velocity).
\\
Using properties of these functions 
the solution of the principal equation is 
\begin{eqnarray} 
	\dfrac{-\lambda^2\vvarpi^2}{k^2v_{\rm A}^2-\omega^2} 
	\dfrac{1}{\rho_0Y}=  
	\dfrac{-\lambda^2k^2c_s^2\Omega^2 \vvarpi^2}{(k^2v_{\rm A}^2-\omega^2)(k^2c_s^2-\omega^2)}
	+\dfrac{a}{2}\dfrac{\vvarpi^2}{\rin^2}-{\rm z}
	\nonumber \\
	+2\dfrac{ (\upalpha-2)M(\upalpha-1,2,{\rm z})+C U(\upalpha-1,2,{\rm z})}{M(\upalpha,2,{\rm z})+CU(\upalpha,2,{\rm z})}-2(\upalpha-1) 
	\,,  
	\label{solprinccomprconstomega}
\end{eqnarray} 
where $C=C_2/C_1$.

We can use the above solution to obtain an analytic expression for the dispersion relation of axisymmetric disturbances in rotating flows in contact, by applying the appropriate boundary conditions at the two extreme radii. Consider, for example, a configuration where the rotating flow occupies $0\le\vvarpi<\rin$ and is surrounded by a static, homogeneous environment.
This setup is analogous to the case discussed in Section~\ref{secincomprtwoflows} for ${\cal A}=1$ (zero rotation in the outer region), but here compressibility effects are included.  
The characteristics of the outer part (density, sound and Alfv\'en velocities) are denoted with the subscript 2, with pressure balance implying $\dfrac{\rho_2c_{s2}^2}{\Gamma_2}+\dfrac{\rho_2v_{\rm A 2}^2}{2}=
\dfrac{\rhoin c_s^2}{\Gamma}+\dfrac{\rhoin v_{\rm A}^2}{2}$. 
\\
The requirement that the solution be smooth on the axis implies
that in the rotating regime $\vvarpi<\rin $ the solution of the principal equation is given by equation~(\ref{solprinccomprconstomega}) with $C=0$. 
The requirement to match the exterior solution given by equation~(\ref{boundaryinftybesselK}) at $\vvarpi=\rin$
(i.e., enforcing continuity of $Y$) yields the dispersion relation 
\begin{eqnarray} 
	\dfrac{\rhoin\Omega^2\rin}{1-\omega^2/k^2c_s^2}=
	\rhoin (k^2v_{\rm A}^2-\omega^2)
	\dfrac{\dfrac{a}{2}-\zin
		+2\dfrac{ (\upalpha-2)M(\upalpha-1,2,\zin) }{M(\upalpha,2,\zin) }-2(\upalpha-1) 
	}{ \lambda^2\rin} 
	\nonumber  \\
+\dfrac{\rho_2}{-i\lambda_2}
(k^2v_{\rm A 2}^2-\omega^2) \dfrac{K_0(-i\lambda_2\rin)}{K_1(-i\lambda_2\rin)}  
\,, 
\label{dispkummer}
	\\ \mbox{where } \quad 
	\lambda^2=-\dfrac{(k^2v_{\rm A}^2-\omega^2)(k^2c_s^2-\omega^2)}{k^2c_s^2v_{\rm A}^2-(c_s^2+v_{\rm A}^2)\omega^2}\,, \quad \Lambda=\sqrt{ k^2 -\dfrac{4\omega^2 \Omega^2k^2  }{(k^2v_{\rm A}^2-\omega^2)^2} }
	\,,
\nonumber 	\\
	a=\dfrac{\rin^2\Omega^2}{c_s^2/\Gamma+v_{\rm A}^2/2}
	\,, \quad 
	a'
	= \dfrac{-\lambda^2k^2 a^2\left[2(1-1/\Gamma)c_s^2-v_{\rm A}^2\right]
		\left(2c_s^2/\Gamma+v_{\rm A}^2\right)}{4(k^2v_{\rm A}^2-\omega^2)(k^2c_s^2-\omega^2)} 
	\,,  
	\nonumber
	\\
	\zin=\sqrt{\dfrac{a^2}{4}+a'} \,, \quad 
	\upalpha=1-\dfrac{ \lambda^2 \Lambda^2\rin^2}{4k^2\zin} 
	\,, \quad
	-i\lambda_2 =\sqrt{\dfrac{(k^2v_{\rm A 2}^2-\omega^2)(k^2c_{s 2}^2-\omega^2)}{k^2c_{s 2}^2v_{\rm A 2}^2-(v_{\rm A 2}^2+c_{s 2}^2)\omega^2}} 
		\,. \nonumber 
\end{eqnarray}

\section{Analytical Solution for Rotating Magnetized Flow with Constant $V_{0\phi}$}\label{appendixbesel}

We can analytically solve the principal equation~(\ref{eqprincipalrotshell}) for axisymmetric disturbances in the case $\Omega =\dfrac{V_0}{\vvarpi}$ with constant $V_0$, assuming also that the sound and Alfv\'en velocities are constant.
The equilibrium condition of the unperturbed state $\Pi_0'= \rho_0\Omega^2 \vvarpi$,
writing $\Pi_0=\dfrac{\rho_0V_0^2}{a} $ using the definitions of the sound and Alfv\'en velocities
($c_s^2= {\Gamma P_0}/{\rho_0}$, $v_{\rm A}^2= {B_0^2}/{\rho_0}$)
and defining the parameter $a=\dfrac{V_0^2}{c_s^2/\Gamma+v_{\rm A}^2/2}$,
yields the radial dependence of the density $\rho_0= \rhoin\left(\dfrac{\vvarpi}{\rin}\right)^a$, and consequently of all the other equilibrium quantities 
$\Pi_0=\dfrac{\rho_0V_0^2}{a}$,
$P_0=\dfrac{\rho_0c_s^2}{\Gamma}$, 
$\bm B_0=v_{\rm A}\sqrt{\rho_0}\hat z$, $\bm V_0=V_0\hat\phi$.  

The analytical solution is
\begin{eqnarray} \dfrac{1}{Y}=\dfrac{\rho_0(\omega^2-k^2v_{\rm A}^2) }{\lambda \vvarpi }\dfrac{ J_{N+1}(\lambda\vvarpi)+C Y_{N+1}(\lambda\vvarpi)}{ J_N(\lambda\vvarpi)+C Y_N(\lambda\vvarpi)}
	\nonumber	\\
	-\left(N+1-\dfrac{a}{2}\right)\dfrac{\rho_0(\omega^2-k^2v_{\rm A}^2) }{\lambda^2\vvarpi^2 }
	-\dfrac{\rho_0k^2V_0^2c_s^2}{ (\omega^2-k^2c_s^2)\vvarpi^2} 
	\,, \label{sol1overY}
	\\ 
\lambda^2=-\dfrac{(k^2v_{\rm A}^2-\omega^2)(k^2c_s^2-\omega^2)}{k^2c_s^2v_{\rm A}^2-(c_s^2+v_{\rm A}^2)\omega^2}
	\,, 
	\\ 
	N^2=\left(\dfrac{a}{2}-1\right)^2 -\dfrac{k^2V_0^4 +4 \omega^2 V_0^2\dfrac{\omega^2-k^2c_s^2}{\omega^2-k^2v_{\rm A}^2} 
		-(a-2)k^2V_0^2c_s^2 
	}{ k^2c_s^2v_{\rm A}^2-(c_s^2+v_{\rm A}^2)\omega^2 }
	\,. 
\end{eqnarray} 

One way to obtain this solution is to cast the principal equation into the associated Sturm-Liouville form~(\ref{odefory1})
\begin{eqnarray}
	\vvarpi^2\xxx''+(a-1)\vvarpi \xxx'+\lambda^2\vvarpi^2\xxx =
	\nonumber	\\ 
	\left[\dfrac{k^2V_0^4 \left(\omega^2- k^2 v_{\rm A}^2\right) +4 \omega^2 V_0^2(\omega^2-k^2c_s^2) -(a-2)k^2V_0^2c_s^2(\omega^2-k^2v_{\rm A}^2)
	}{(\omega^2-k^2v_{\rm A}^2)^2(\omega^2-k^2c_s^2) }
	\right]\lambda^2\xxx
	\,,  
	\\
	\dfrac{1}{Y}=-\dfrac{\rho_0(\omega^2-k^2v_{\rm A}^2)\xxx'}{\lambda^2\vvarpi \xxx}
	-\dfrac{\rho_0k^2V_0^2c_s^2}{ (\omega^2-k^2c_s^2)\vvarpi^2}
	\,, 
\end{eqnarray}
which is transformed to Bessel $\vvarpi^2 B''+\vvarpi B'+(\lambda^2\vvarpi^2-N^2)B=0$
with the substitution $\xxx=\vvarpi^{1-a/2}B(\vvarpi)$.
The general solution is $\xxx=C_1\vvarpi^{1-a/2}J_N(\lambda\vvarpi)+C_2\vvarpi^{1-a/2}Y_N(\lambda\vvarpi)$
giving the expression~(\ref{sol1overY}) with $C=C_2/C_1$.

For axisymmetric disturbances of the rotating magnetized shell considered in Ref.~\cite{2019MNRAS.488.4061K}, we can use the above solution for the interior of the shell and obtain analytic expression of the dispersion relation by applying the appropriate boundary conditions at its two extreme radii. 
\\
The requirement at the inner wall $Y|_{\rin-\Delta\vvarpi}=0$ gives $C=-\dfrac{J_N(\lambda\rin-\lambda\Delta\vvarpi)}{Y_N(\lambda\rin-\lambda\Delta\vvarpi)}$, and the requirement to match the exterior solution given by equation~(\ref{boundaryinftybesselK}) at $\vvarpi=\rin$ yields the dispersion relation 
\begin{eqnarray} 
N=\dfrac{a}{2}-1+\lambda \rin \dfrac{ Y_N(\lambda\rin-\lambda\Delta\vvarpi) J_{N+1}(\lambda\rin)- J_N(\lambda\rin-\lambda\Delta\vvarpi) Y_{N+1}(\lambda\rin)}{  Y_N(\lambda\rin-\lambda\Delta\vvarpi) J_N(\lambda\rin)- J_N(\lambda\rin-\lambda\Delta\vvarpi) Y_N(\lambda\rin)}
\nonumber \\ 
-\dfrac{k^2V_0^2c_s^2}{ (c_s^2+v_{\rm A}^2)\omega^2-k^2c_s^2v_{\rm A}^2 }
+\dfrac{\rho_2}{\rhoin}\dfrac{\omega^2-k^2v_{\rm A 2}^2}{\omega^2-k^2v_{\rm A}^2}
\dfrac{\lambda^2\rin^2 }{-i\lambda_2\rin}\dfrac{ K_0(-i\lambda_2\rin)}{K_1(-i\lambda_2\rin)}  
	\,, 
	\\ \mbox{where} \quad 
	\lambda=\sqrt{\dfrac{(\omega^2-k^2v_{\rm A}^2)(\omega^2-k^2c_s^2)}{(c_s^2+v_{\rm A}^2)\omega^2-k^2c_s^2v_{\rm A}^2}}
	\,, \quad  
	-i\lambda_2 =\sqrt{\dfrac{(k^2v_{\rm A 2}^2-\omega^2)(k^2c_{s 2}^2-\omega^2)}{k^2c_{s 2}^2v_{\rm A 2}^2-(v_{\rm A 2}^2+c_{s 2}^2)\omega^2}}
	\,. \nonumber
	\end{eqnarray}  

\begin{adjustwidth}{-\extralength}{0cm}
\printendnotes[custom] 

\reftitle{References}

\vspace{-1mm}
\PublishersNote{}
\end{adjustwidth}
\end{document}